\documentclass{article}

\usepackage[nonatbib,final]{neurips_2021}

\usepackage[backend=biber,style=numeric,sorting=none,maxbibnames=3,doi=false,isbn=false,url=true,eprint=false]{biblatex}
\AtEveryBibitem{\clearlist{language}}
\usepackage[utf8]{inputenc} 
\usepackage[T1]{fontenc}    
\usepackage{url}            
\usepackage{booktabs}       
\usepackage{amsfonts}       
\usepackage{nicefrac}       
\usepackage{microtype}      
\usepackage{xcolor}         
\usepackage{wrapfig}

\usepackage{float}
\usepackage{makecell}
\usepackage{multirow}
\usepackage{subcaption}
\usepackage[colorlinks=true, linkcolor=blue, citecolor=red, linktocpage=true]{hyperref}

\usepackage{graphicx}
\graphicspath{ {./}
	{./pics/} 
	{./output/motifs/pics/} 
	{./output/benchmark/} 
	{./output/lime/}
	{./output/shap/pics/} 
	{./output/shap/pics/change/}  
}

\usepackage[acronym]{glossaries}
\newacronym{MHC}{MHC}{major histocompatibility complex}
\newacronym{MHC-I}{MHC-I}{MHC class I}
\newacronym{MHC-II}{MHC-II}{MHC class II}
\newacronym{SOTA}{SOTA}{state of the art}
\newacronym{BERT}{BERT}{Bidirectional Encoder Representations for Transformers}
\newacronym{SHAP}{SHAP}{SHapley Additive exPlanations}
\newacronym{LIME}{LIME}{Local Interpretable Model-agnostic Explanations}
\newacronym{CTL}{CTL}{Cytotoxic T-lymphocytes}
\newacronym{HLA}{HLA}{Human Leukocyte Antigen}
\newacronym{TCR}{TCR}{T-cell receptor}
\newacronym[plural=pMHCs,firstplural=peptide:MHC protein complexes (pMHCs)]{pMHC}{pMHC}{peptide:MHC protein complex}
\newacronym{AA}{AA}{amino acid}
\newacronym{SA}{SA}{single-allele}
\newacronym{MA}{MA}{multi-allele}
\newacronym{DNA}{DNA}{deoxyribonucleic acid}
\newacronym{mRNA}{mRNA}{messenger ribonucleic acid}
\newacronym{TAP}{TAP}{transporter associated with antigen processing}
\newacronym{ER}{ER}{endoplasmic reticulum}
\newacronym{EL}{EL}{eluted ligand}
\newacronym{BA}{BA}{binding affinity}
\newacronym{MIL}{MIL}{Multi Instance Learning}
\newacronym[plural=PWMs,firstplural=position weight matrices (PWMs)]{PWM}{PWM}{position weight matrix}
\newacronym{PU}{PU}{positive and unlabelled}
\newacronym{AUC}{AUC}{area-under-the-curve}
\newacronym{TP}{TP}{true positive}
\newacronym{FP}{FP}{false positive}
\newacronym{TN}{TN}{true negative}
\newacronym{FN}{FN}{false negative}
\newacronym{TPR}{TPR}{true positive rate}
\newacronym{FPR}{FPR}{false positive rate}
\newacronym{AP}{AP}{average precision}
\newacronym{PR}{PR}{precision-recall}
\newacronym{MLM}{MLM}{Masked Language Modelling}
\newacronym{NSP}{NSP}{Next Sentence Prediction}
\newacronym{TAPE}{TAPE}{Tasks Assessing Protein Embeddings}
\newacronym{ML}{ML}{machine learning}
\newacronym{IC}{IC}{information content}
\newacronym{LSTM}{LSTM}{Long Short-term Memory}
\newacronym{ResNet}{ResNet}{residual network}
\newacronym{MLP}{MLP}{multi layer perceptron}
\newacronym{ROC}{ROC}{receiver-operating-curve}
\newacronym{CNN}{CNN}{convolutional neural network}
\newacronym{PRIDE}{PRIDE}{PRoteomics IDEntifications Database}
\makeglossaries

\newcommand{\finalModel}{epoch=4-step=3648186}
\newcommand{\HLAalleleI}{HLA-A3301}
\newcommand{\HLAalleleIname}{HLA-A*33:01}
\newcommand{\HLAalleleII}{HLA-B5401}
\newcommand{\HLAalleleIIname}{HLA-B*54:01}
\newcommand{\HLAalleleIII}{HLA-C0102}
\newcommand{\HLAalleleIIIname}{HLA-C*01:02}
\newcommand{\HLAallele}{HLA-A3301}
\newcommand{\HLAalleleName}{HLA-A*33:01}

\title{Interpreting BERT architecture predictions for peptide presentation by MHC~class~I proteins}

\author{%
  Hans-Christof Gasser \\
  School of Informatics\\
  University of Edinburgh\\
  \texttt{h.gasser@sms.ed.ac.uk} \\
  \And
  Georges Bedran \\
  ICCVS\\
  University of Gdańsk\\
  \texttt{georges.bedran@phdstud.ug.edu.pl} \\
  \And
  Bo Ren \\
  Biochemistry and Microbiology \\
  University of Victoria\\
  \texttt{boren@uvic.ca}\\
  \And
  David Goodlett \\
  Biochemistry, Microbiology \\
  and GBC Proteome Centre\\
  University of Victoria\\
  \texttt{goodlett@uvic.ca}\\
  \And
  Javier Alfaro \thanks{jointly supervised} \\
  ICCVS\\
  University of Gdańsk\\
  \texttt{javier.alfaro@ug.edu.pl} \\
  \And
  Ajitha Rajan {\scriptsize *} \\
  School of Informatics\\
  University of Edinburgh\\
  \texttt{arajan@ed.ac.uk} \\
}

\begin{document}

\maketitle

\begin{abstract}
The \gls{MHC} class-I pathway supports the detection of cancer and viruses by the immune system. It presents parts of proteins (peptides) from inside a cell on its membrane surface enabling visiting immune cells that detect non-self peptides to terminate the cell. The ability to predict whether a peptide will get presented on MHC Class I molecules helps in designing vaccines so they can activate the immune system to destroy the invading disease protein. We designed a prediction model using a BERT-based architecture (ImmunoBERT) that takes as input a peptide and its surrounding regions (N and C-terminals) along with a set of \gls{MHC-I} molecules. We present a novel application of well known interpretability techniques, \acrshort{SHAP} and \acrshort{LIME},  to this domain and we use these results along with 3D structure visualizations and amino acid frequencies to understand and identify the most influential parts of the input amino acid sequences contributing to the output. In particular, we find that amino acids close to the peptides' N- and C-terminals are highly relevant. Additionally, some positions within the \gls{MHC} proteins (in particular in the A, B and F pockets) are often assigned a high importance ranking - which confirms biological studies and the distances in the  structure visualizations. The source code can be found on \url{https://github.com/hcgasser/ImmunoBERT}.
\end{abstract}

\section{Introduction}
\begin{figure}
	\centering
	\begin{subfigure}{0.4\textwidth}
			\centering
			\includegraphics[width=51mm]{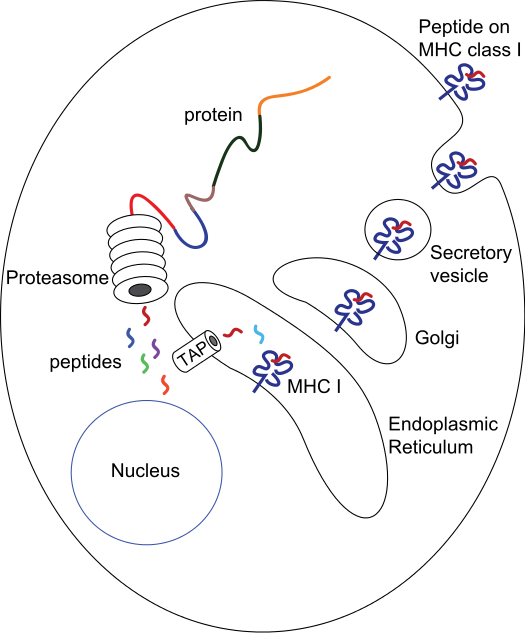}
			\caption[MHCpathway]{\gls{MHC-I} pathway \cite{wikipedia_major_2021}}
			\label{MHCpathway}
	\end{subfigure} 
	\hfill
	\begin{subfigure}{0.55\textwidth}
		{\centering
	 	\includegraphics[width=0.55\textwidth]{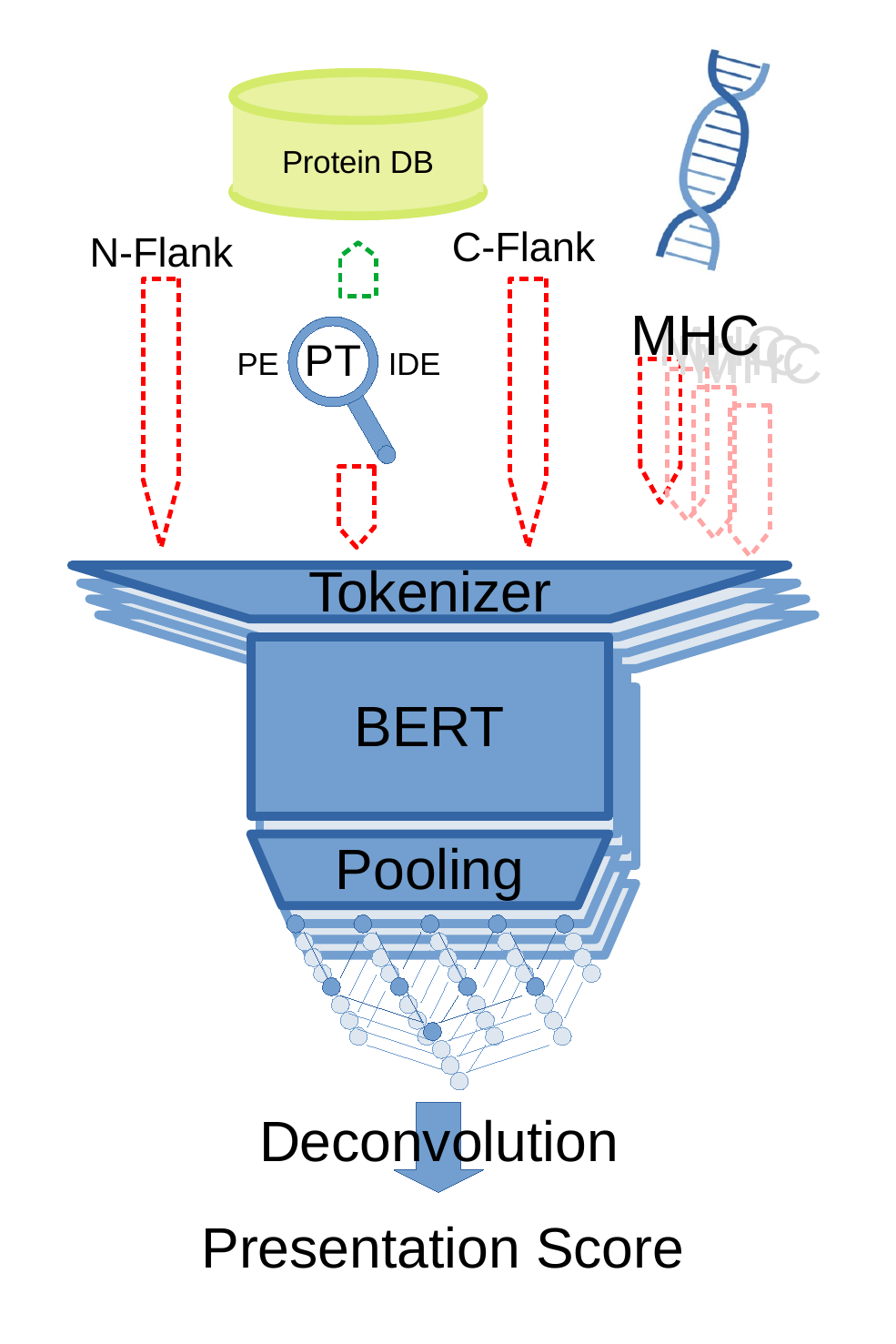}
		\caption{ImmunoBERT for peptide presentation on MHC-I}	
	 	\label{fig:summary}
		}
	\end{subfigure}
	\caption{MHC-I pathway and our prediction model for peptide presentation}
	\label{Fig:Intro}
	\end{figure}

The immune system defends us from a broad range of threats, some of which are expressed from inside the body's own cells. For example cancer is a disease of the genome, arising from aberrations that accumulate over many years. Also, viruses utilize the cell's gene expression system for their own reproduction and spreading. \gls{CTL}, a special kind of T-cells, can detect affected cells and terminate them. For them to `look inside' cells, a system revolving around \gls{MHC} proteins - also called \gls{HLA} in humans - has evolved \cite{encyclopaedia_britannica_major_2021}. We focus on \gls{MHC-I} proteins and their antigen presentation pathway which is active in the nucleated cells of the human body \cite{mosaad_clinical_2015}. Proteins present in these cells are constantly being fragmented into smaller pieces (peptides) by proteasomes (see Figure \ref{MHCpathway}). These then bind to \gls{MHC} proteins forming \glspl{pMHC} and are eventually presented to the outside world on the cell's surface. These \gls{pMHC} are antigens for the \glspl{TCR}. \cite{murphy_janeways_2017}

An infection by a virus or a cancer causing mutation, can both result in the production of proteins that would not be present in a healthy cell. Eventually this leads to the presentation of neo-antigens (non-self \gls{pMHC}) to the outside world \cite{schumacher_neoantigens_2015, zhang_neoantigen_2021}. Dependent on whether the \gls{CTL} consider the peptides presented to them as self or non-self, they decide to terminate the cell. Peptide-based vaccines are a tool that can be used to strengthen the immune response. Identifying which non-self peptides will most likely be presented by a cancerous cell or virus and elicit an immune response (immunogenic peptides \cite{li_deepimmuno_2021}) is an important component in their development \cite{comber_mhc_2014}. This process can be divided into two phases -- (1) Identification of peptides that are likely to be presented by the \gls{MHC} protein which is the focus of this paper, (2) Predicting activation of the immune system by the presented peptides (outside the paper's scope).
Deep learning models have been used for Phase 1 that identifies presented peptides on MHC molecules. 
\citeauthor{cheng_bertmhc_2020} used a bi-directional transformer architecture to predict peptide presentation by a related molecule - \gls{MHC-II}. We adapted this work for the \gls{MHC-I} presentation prediction and find our model, ImmunoBERT, provides comparable performance to popular \gls{SOTA} models, NetMHCpan \cite{reynisson_netmhcpan-41_2020} and MHCflurry \cite{odonnell_mhcflurry_2020} (see Appendix Subsection \ref{subsec:results_benchmarking}).
ImmunoBERT, as seen in Figure \ref{fig:summary},  takes a peptide, its surrounding regions (N-flank and C-flank) and \gls{MHC-I} molecules as inputs  and provides as output a score that represents the likelihood of the peptide being presented on \gls{MHC-I} molecules. 

In this paper, we focus on interpreting the results from our deep learning model. We apply \gls{LIME} \cite{ribeiro_why_2016} and \gls{SHAP} \cite{lundberg_unified_2017} to find the parts of the peptide, \gls{MHC} protein and surrounding flanks of the peptide, that are particularly relevant for presentation. We corroborate the results of \gls{LIME} and \gls{SHAP} using visualizations that present 3D peptide-\gls{MHC-II} molecule interactions and a motif (short, recurring patterns \cite{dhaeseleer_what_2006}) analysis that confirms that our model has indeed learnt \gls{MHC} allele dependent peptide presentations.

Section \ref{section:background} below introduces the necessary biological concepts, while Section \ref{sec:related_work} presents the current \gls{SOTA} approaches. In Section \ref{section:method}, we introduce the datasets and our ImmunoBERT model. Finally, we present model interpretation and visualization in Section \ref{section:results}.

\section{Background}
\label{section:background}

    Proteins play a pivotal role in the human body. They consist of \glspl{AA} chained together via peptide bonds into a polypeptide or peptide for short \cite{pollard_cell_2016}. There are 20 naturally occurring amino acids and each contains a specific side chain that gives it unique properties to interact with the environment \cite{crowe_chemistry_2014, pollard_cell_2016}. The end of the chain with an exposed amino group is called N-terminus, the end with an exposed carboxyl group is called C-terminus. Each of them can be represented by a letter. Once joined together, we refer to the single \glspl{AA} as residues. Short residue chains are referred to as peptides, while longer ones are referred to as polypeptides. The full-length polypeptide encoded by a gene in the genome is called a protein. Proteins typically adopt a 3D molecule structure due to the chemical properties of their \glspl{AA}.

	Proteins perform diverse tasks like breaking up nutrition, muscle movement and sustaining cell structure. Owing to their wide range of functionality, and for protection against foreign invading proteins from viruses and cancers, controlling which proteins are present in a cell is important. 
	The \gls{MHC-I} pathway achieves this (see Figure \ref{MHCpathway}) \cite{rock_present_2016} with the following steps:
	
		\begin{enumerate}
			\item Proteasomes constantly fragment the cell's internal proteins into peptides (mostly 8-12 \glspl{AA} long \cite{nielsen_immunoinformatics_2020}). For example, part of the \gls{AA} sequence of the protein Albumin is  \texttt{MKWVTFISLLFLFSS\textcolor{orange}{AYSRGVFRRDAHKSE}\textcolor{green}{VAHRFKDLG}\textcolor{orange}{EENFKALVLIAFAQY}LQQCP..} and the proteasomes could fragment this into peptides -- "\texttt{VTFISLLFL}", "\texttt{\textcolor{green}{VAHRFKDLG}}" and "\texttt{FAQYLQQ}". Each of these peptides are associated with a N-flank and a C-flank. For example, the peptide "\texttt{\textcolor{green}{VAHRFKDLG}}" would have a 15 \gls{AA} N-flank of "\texttt{\textcolor{orange}{AYSRGVFRRDAHKSE}}" and a 15 \gls{AA} C-flank of "\texttt{\textcolor{orange}{EENFKALVLIAFAQY}}".

			\item \Gls{TAP} proteins transport these peptides into the \gls{ER} 
			where they bind to \gls{MHC-I} proteins, forming \gls{pMHC} with the peptides.

			\item The \glspl{pMHC} gets transported to the cell membrane, where the \gls{MHC} protein acts as a pedestal for the peptide and presents it to the extracellular environment 
			\item Finally,  A \gls{CTL} with a fitting \gls{TCR} could bind a presented neo-antigen. This might trigger an immune reaction. \glspl{CTL} do not strongly bind to the body's own (self) peptides but only non-self ones. 
		\end{enumerate}
	We focus on steps 1-3 in this paper that form Phase 1 of identifying non-self peptides by the immune system.	There are many different \gls{HLA} alleles \cite{pollard_cell_2016, murphy_janeways_2017} in the human population with three main loci coding for \gls{MHC-I} proteins: HLA-A, HLA-B and HLA-C \cite{mosaad_clinical_2015}, with every human expressing up to six different HLA proteins.  
	As the HLA alleles can have different binding properties there is a large variety in immunopeptidomes (entirety of all presented peptides) across humanity. Currently, more than  4,064 HLA-A, 4,962 HLA-B and 3,831 HLA-C proteins are known \cite{ebi_immuno_2021}. Each can bind roughly 1,000 to 10,000 different peptides \cite{abelin_mass_2017}.
	
    Predicting the immunopeptidome of a particular individual is challenging owing to two reasons. First, any of the six different \gls{MHC-I} alleles present in a cell might be responsible for a peptide observation in an \gls{EL} experiment (see Appendix Subsection \ref{subsec:experimental_datasources}). So the observations need to be deconvoluted (assigned to a HLA allele). Second, the high throughput eluted ligand assays only produce positive examples resulting in a highly imbalanced dataset that requires the creation of artificial negative ones (decoys). This is exacerbated by the fact that available labelled data accounts for a very small fraction of the peptides that can/cannot be presented. Absence of ground truth, variance among individuals, limited labeled data makes prediction modeling in the field of immunopeptidomics extremely challenging.     

\section{Related Work}
\label{sec:related_work}

    Below, we take a brief look at the two most popular state of the art prediction models for peptide presentation on MHC-I molecules.
    
	\textbf{NetMHCpan: }
	The most commonly used model today is NetMHCpan. It has a long history and is currently in version 4.1 which is an ensemble of 50 single hidden layer feed forward neural networks \cite{reynisson_netmhcpan-41_2020}. The \gls{MHC} allele is input into the model as a pseudo sequence consisting of only 34 \glspl{AA}. These were identified by \cite{nielsen_netmhcpan_2007} as being particularly close to the presented peptide and relevant for peptide binding.

	To deconvolute \gls{MA} data, NetMHCpan uses the NNAlign\_MA \cite{alvarez_nnalign_ma_2019} framework. First, only \gls{SA} data (the observation can be unambigiously linked to a single \gls{MHC} protein) is used to train a classifier (takes as input a peptide and a single \gls{MHC} allele). In the deconvolution step,  each observation that could be caused by multiple \gls{MHC} alleles, is deconvolved separately. To do so, the classifier trained in the previous step is used to predict the likelihood of each potential peptide:MHC protein combination independently. The \gls{MHC} allele showing the highest scaled prediction is chosen and used as the MHC protein responsible for the observation until the next deconvolution step. In case of a negative example, a \gls{MHC} allele is picked at random.

	\textbf{MHCflurry 2.0} \cite{odonnell_mhcflurry_2020} explicitly models the process of \gls{MHC} binding separately from the others (e.g. proteasomal cleavage). This results in a natural integration of \gls{BA} and \gls{EL} data (see Appendix Subsection \ref{subsec:experimental_datasources}). There are three sub-models. First, MHCflurry BA models the process of the peptide binding to a \gls{MHC} protein. Second, MHCflurry AP models the remaining antigen processing steps, like proteasomal cleavage and \gls{TAP} transportation. Finally, MHCflurry PS combines the output of those two models to predict peptide presentation. \citeauthor{odonnell_mhcflurry_2020} \cite{odonnell_mhcflurry_2020} benchmarked their performance on held-out MS data against NetMHCpan 4.0 and  MixMHCpred 2.0.2. They found their model had better performance (with regards to their chosen metric - positive predictive value). We use this benchmark dataset for comparing our model, ImmunoBERT, against \gls{SOTA} models, NetMHCpan and MHCflurry 2.0 (see Appendix Subsection \ref{subsec:results_benchmarking}). 
	
	Some work has been done to interpret protein data predictions. For example \citeauthor{vig_bertology_2021} \cite{vig_bertology_2021} have used the transformers attention mechanism to show that some of the transformer's nodes were able to learn biological properties of proteins (e.g. secondary structure, binding sites, ...). However, these attention based approaches are hardly suitable for the explanation of a single prediction and to our knowledge, there is no existing work on interpreting deep learning models for peptide presentation with  MHC molecules. However, several interpretation techniques for deep learning models have been recently developed in the field of computer vision - popular ones include \gls{LIME} \cite{ribeiro_why_2016} and \gls{SHAP} \cite{lundberg_unified_2017}. \gls{LIME} produces local (for a particular example) explanations, treating the model to be explained as a black-box (model-agnostic). Given a particular input, \gls{LIME} samples the
neighborhood of this input and creates a linear model to approximate the model’s local behavior. In comparison, \gls{SHAP} values are based on the idea of Shapley values, that attribute the difference between the average prediction over the dataset and the example's prediction fairly to the various features. \citeauthor{lundberg_unified_2017} \cite{lundberg_unified_2017} developed the \gls{SHAP} package for the efficient approximation of \gls{SHAP} values under the assumption that the model's output restricted to a subset of the features is given by the expected model prediction conditioned on this subset. We apply these two techniques to the problem of peptide presentation and combine it with visualizations and anlysis accessible to biologists and clinicians. 

\section{Method}
\label{section:method}
In this section, we present the peptide and MHC-I data used to train and validate ImmunoBERT, the model architecture and interpretation techniques. Source code for our model and interpretation can be found at \url{https://github.com/hcgasser/ImmunoBERT}.

	\subsection{Data}
	\label{subsec:Data}
		We combined data from two sources. The first one consists of a collection of peptides from \gls{EL} assays mapped to the GRCh38 \textit{Homo sapiens} reference genome and proteins within the Ensembl v94 database. The source of this data is studies included in the \gls{PRIDE} \cite{perez-riverol_pride_2019}. We removed samples without linked \gls{HLA} proteins or where the peptide is not present in the human proteome (neo-antigens). The second data source is the HLA Ligand Atlas \cite{marcu_hla_2020}. This includes tissue and \gls{HLA} allele specific ligands from \gls{EL} experiments. In contrast to the first data source, the HLA Ligand Atlas maps peptides to the Uniprot proteome, some of which map to GRCh38. Similar to other studies, we consider peptides of sequence lengths between 7 and 15 amino acids~\cite{reynisson_netmhcpan-41_2020, odonnell_mhcflurry_2020}.
		
		A sample represents the experiment carried out on a particular cell-line/individual to measure the presented peptides. It is, therefore, linked with up to 6 different \gls{HLA} alleles and the observations (also referred to as \emph{hits}) of peptides during the experiment. Each peptide was mapped to proteins in the human genome. 
		In total we had 293k \gls{SA} and 1,666k \gls{MA} observations of 430k unique peptides. These were observed in 469 samples linked with 109 different \gls{MHC} alleles.
			
		\textbf{Decoy generation: }	Negative example (decoy) generation is particularly important due to the imbalanced nature of the dataset. We follow a similar procedure to the MHCflurry benchmark dataset\cite{odonnell_mhcflurry_2020}. A decoy is associated with a single hit. To match the observations' length distribution, the decoy peptides have the same length as their associated hit. To generate a decoy peptide we randomly select a position within all proteins of the hit's sample as the start of the decoy peptide. Implicitly we take the absence of a peptide's observation as evidence for it actually not being presented.  We considered using 19 or 99 decoys per hit in our hyper-parameter search.
		
		\textbf{Data splits: } Splitting the data into train, test and validation set is not trivial, as we assessed generalization along 2 dimensions - to unseen \gls{MHC} alleles and unseen proteins. Also, each observation can be associated with up to six \gls{MHC} alleles from which at least one is responsible for the presentation. There are also many homologues (areas having a common ancestor) in the human genome and, ideally, a group of homologues would not span different splits. Our methodology for splitting the data can be found in Appendix Subsection \ref{subsec:data_split}. We get one training set, two validation sets and two test sets. The data partition can also be found in the Appendix.
	
	\subsection{Model Architecture and Training} 
	    We use the pre-trained \gls{TAPE} transformer as backbone for our model (a BERT architecture adapted to amino acid sequences). As input we provide our model with the peptide's \gls{AA} sequence. If uniquely identified, we also input the peptide's context which consists of (1) up to 15 \glspl{AA} that occur to the \emph{left} of the peptide in its source protein sequence known as \emph{N-flank}, and (2) up to 15 \glspl{AA} that occur to the \emph{right} referred to as \emph{C-flank}. Finally, the model also receives the \gls{MHC} pseudo sequence as defined by NetMHCpan (see Subsection \ref{sec:related_work}).
		
		Off the shelf, the \gls{TAPE} transformer supports only a single token type id. To make it easy to distinguish between the various input parts, we use a novel representation of the input and extend the \gls{TAPE} model's token type embedding matrix to four different token types - one for the N-flank, peptide, C-flank and \gls{MHC} protein. The resulting embedding vectors are fed through the \gls{TAPE} encoder (12 self attention layers with 12 heads each). In contrast to BERTMHC \cite{cheng_bertmhc_2020} that then uses average pooling with the embedded vectors, we explored three options as part of the hyper-parameter search: averaging, attention layer or the classification token's vector. This was done to consider the meaning of the peptide, context and \gls{MHC} sequence.  After comparison of the three options, we chose the classification token's vector that was best performing.
			
		The structure of our model's head is similar to the one used by BERTMHC \cite{cheng_bertmhc_2020}. It is also a \gls{MLP} consisting of two fully connected layers with a hidden dimension of 512. There is a single output neuron, with sigmoid activation for the presentation score.
	    Our model's training procedure is inspired by the  NNAlign\_MA framework and general results from \gls{MIL}. In the first training epoch we only use \gls{SA} data. This is followed by a deconvolution phase. During this, we deconvolve each \gls{MA} observation by at first predicting the presentation score for each potentially responsible allele and then selecting the one with the highest score as the relevant allele. This means, the \gls{MA} example is treated (forward-propagation and backward-propagation) as if coming from the relevant allele until the next deconvolution phase (happens after each epoch). The decoy's relevant \gls{MHC} allele is the same as the observation.
    		
    	For training we use standard binary cross entropy as loss function. We trained the full network (including encoder and embedding layers) using the ADAM optimizer ($\beta_1 = 0.9, \beta_2 = 0.999$ \cite{kingma_adam_2017}). We chose the initial learning rate as  part of the hyper-parameter search.

		\subsubsection{Interpretation}
		\label{subsubsec:method_Interpretation}
			To check, whether our model has learned relationships grounded in biology, we employ \gls{LIME} \cite{ribeiro_why_2016} and \gls{SHAP} \cite{lundberg_unified_2017}. We restrict our analysis to 9-mer peptides and \gls{SA} data. As we interpret test set data, the analysis demonstrates our model's ability to generalize to unseen \gls{MHC} proteins. We additionally use Sequence Motifs and 3D visualization of the MHC and peptide structures to better understand the model results and the interpretations from LIME and SHAP. 

			\noindent \textbf{LIME analysis of all features:} We first interpret the model output by assessing the importance of all \gls{AA} features in the input peptide, context and HLA allele sequences. For this, we use the \gls{LIME} framework. This is conceptually faster than \gls{SHAP} - in particular when handling numerous features (in our case ~73 \gls{AA} positions). \gls{SHAP} could be sped up by using fewer samples for the background distribution or sampling fewer feature subsets. It is then, however, questionable if the results would still be good estimates for \gls{SHAP}'s central claim - to attribute the difference between the unconditional expected prediction and the prediction conditional on the relevant example to the example's particular features. No such strong claim is made by \gls{LIME}.
			
			To use the LIME package for our domain, we adapted the text version approach of \gls{LIME}. This means, we implemented the deactivation of a feature by setting the input mask token to zero. We use the standard cosine distance metric (in binary space) between the original example and the sampled examples. For each test set \gls{HLA} allele, we selected a random set of 500 observations with each of these observations accompanied by a decoy bringing the total to 1000 examples in the test set that we aim to interpret. Each example output gets explained by sampling from 2000 feature combinations. Figures \ref{fig:LIME_\HLAalleleI} and \ref{fig:LIME_\HLAalleleII} then show in each bar the proportion of examples with a given importance-ranking for an \gls{AA} at a certain position. If for example the bright red bar (1st) of peptide position 9 showed 0.5, this means that 50\% of samples in the test set had peptide position 9 as the most important feature.
		
		\noindent \textbf{SHAP analysis of peptide positions:} Finally, we examine the average contribution of peptide \glspl{AA} using Kernel \gls{SHAP} and adapted this to our input structure. Similar to the LIME setup, the interpretation of each \gls{HLA} allele uses 500 9-mer single allele hits and 500 decoys. We sample 250 sequences as background distribution. The nine position features in the peptide would result in a maximum of 512 feature subsets. We carry out the Kernel \gls{SHAP} analysis using a sample of 64 of these. For the whole process we ignore the flanks. In Figures \ref{fig:SHAP_\HLAalleleI} and \ref{fig:SHAP_\HLAalleleII} we plot the average \gls{SHAP} value for each \gls{AA} at each peptide position.	
		
			\noindent \textbf{Sequence Motifs:} We visualize for each \gls{HLA} allele, the frequency of various amino acids at presented peptide positions \cite{schneider_sequence_1990} and contrast this with the feature importance rankings generated by LIME and SHAP. We use two different motifs to understand the results from ImmunoBERT -
			
			\begin{enumerate}
			 \item \emph{Model Motif:} we generate 100,000 random 9-mer peptides from the human proteome (from Uniprot and Ensembl proteins, as well as their context). Then we predict for each of those examples the presentation score for the \gls{HLA} protein concerned. We select the ones with a presentation score $> 0.5$ and use them to create the HLA allele dependent \emph{model motif} (using the logomaker \cite{tareen_logomaker_2019} package). Our approach for creating model motifs is similar to \citeauthor{wu_deephlapan_2019} \cite{wu_deephlapan_2019} who use the 2\%  highest scoring peptides for model motif.
			 \item \emph{Data Motif:} we take the data from our test set and use all of the 9-mer peptides presented by the \gls{HLA} allele to create it.
			\end{enumerate}   
			 
			The data motif shows the frequency of \glspl{AA} at presented peptide positions (independent of our model) based on existing labelled data on peptide presentation within a sample. The model motif, on the other hand, shows the frequency of \glspl{AA} at presented peptide positions predicted by our model - sampled for peptides across the genome. We do not expect both motifs to be the same but if the test set for any given HLA allele were a representative sample from the human proteome, we expect there will be overlap between the data motif and the model motif. 
			
			In a motif logo, for each peptide position a stack of \gls{AA} letters is displayed. The size of each letter is proportional to the \gls{AA}'s frequency at this position. More frequent \glspl{AA} can be found on top. Each stack is then scaled with the position's \gls{IC}, resulting in a bit representation \cite{schneider_sequence_1990}. The lower the position's entropy, the higher the \gls{IC} and, so, the logo. We do not display positions with \gls{IC} $< 0.5$ to avoid distraction. \gls{AA} with similar chemistry are coloured the same. 

            \noindent \textbf{PyMOL visualizations:} PyMOL~\cite{schroedinger_pymol_2021}, a cross-platform molecular graphics tool, has been widely used for three-dimensional (3D) visualization of proteins, nucleic acids, small molecules, electron densities, surfaces, and trajectories. The 3D visualizations presented in this paper are based on the \gls{MHC} 3D structures in \cite{menezes_teles_e_oliveira_phla3d_2019}. We colored them with their mean importance ranking - \gls{AA} that were not used as features are blue, in contrast, the highest important \gls{AA} are colored red. Visualizing the importance of various amino-acids on a structure of peptide bound MHC can help to shed light on how importance correlates with physical interactions known to be important.

\section{Results}
\label{section:results}
	In this chapter, we present the results from applying the interpretability techniques described in \ref{subsubsec:method_Interpretation} to the ImmunoBERT model, limited to peptide presentation by two test set \gls{HLA} alleles. There are many more test \gls{HLA} alleles in our data, but we are unable to fit all their analyses and visualizations within the defined page limit.
	Figures for all the remaining test \gls{HLA} alleles can be found in the Appendix. It is worth noting that to improve readability all \gls{SHAP} values were multiplied by 100.
	
	Visualizing the importance of various amino-acids on a 3D structure of peptide bound MHC can help to shed light on how importance correlates with physical interactions known to be important. The Figures \ref{fig:3D-\HLAalleleI} and \ref{fig:3D-\HLAalleleII} demonstrate that regions near the anchor residues on the MHC molecule are important both within the peptide and the MHC molecule. This re-affirms known biology of antigen binding to MHC and indicates that LIME and SHAP provide reasonable results.

	\subsection{\HLAalleleIname}
\begin{figure}[H]
						\includegraphics[width=14.5cm]{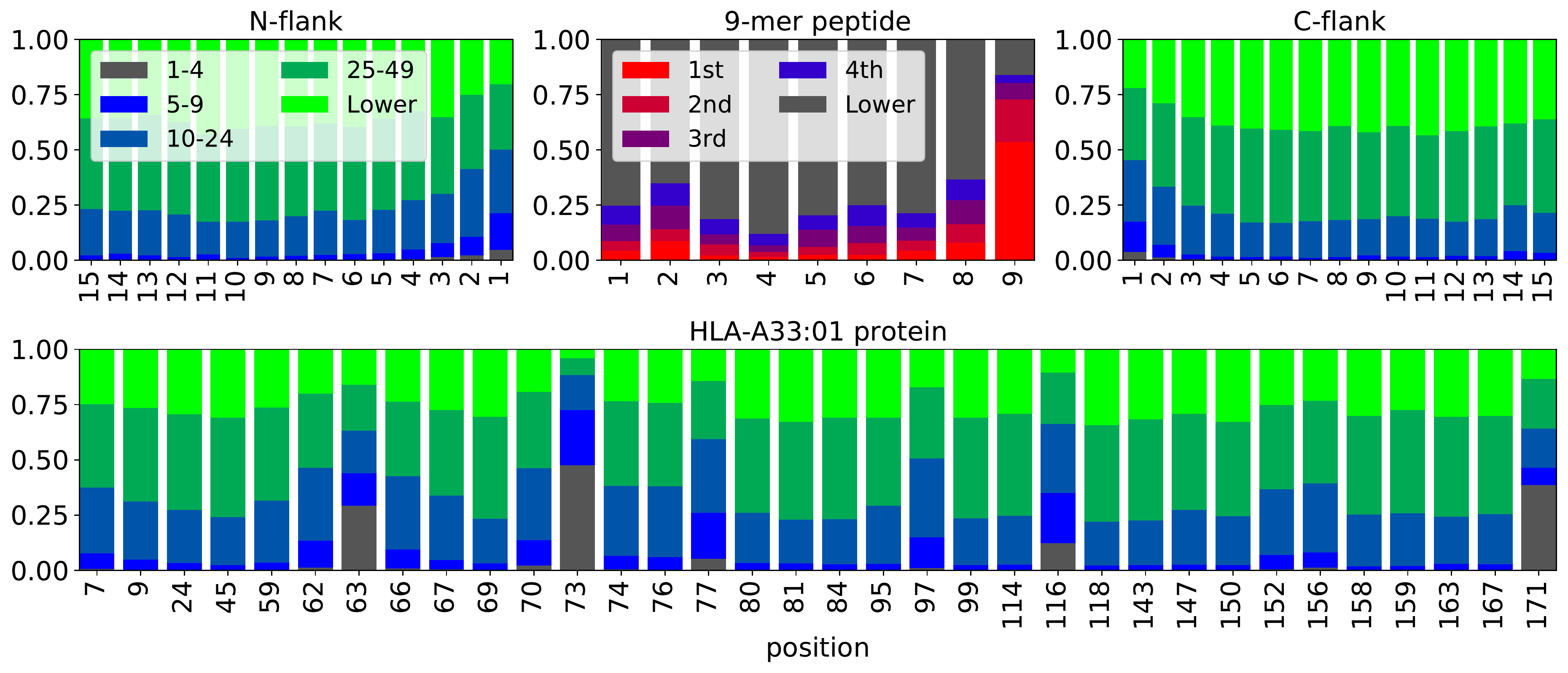} 
						\captionof{figure}{\gls{LIME} feature importance rank distribution for \HLAalleleIname}
						\vspace{-10pt}
						\label{fig:LIME_\HLAalleleI}
				\end{figure}

\begin{figure}
				    \centering
				    \begin{subfigure}[b]{0.45\textwidth}
					\includegraphics[width=6cm]{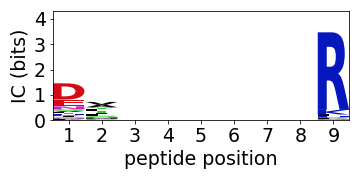}
					\includegraphics[width=6cm]{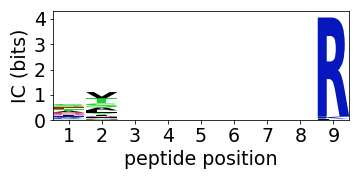} 
					\captionof{figure}{Data(top) \& model motif}
					\label{fig:LOGO_\HLAalleleI}
					\end{subfigure}
                    \hfill
                    \begin{subfigure}[b]{0.45\textwidth}
					\includegraphics[width=6cm]{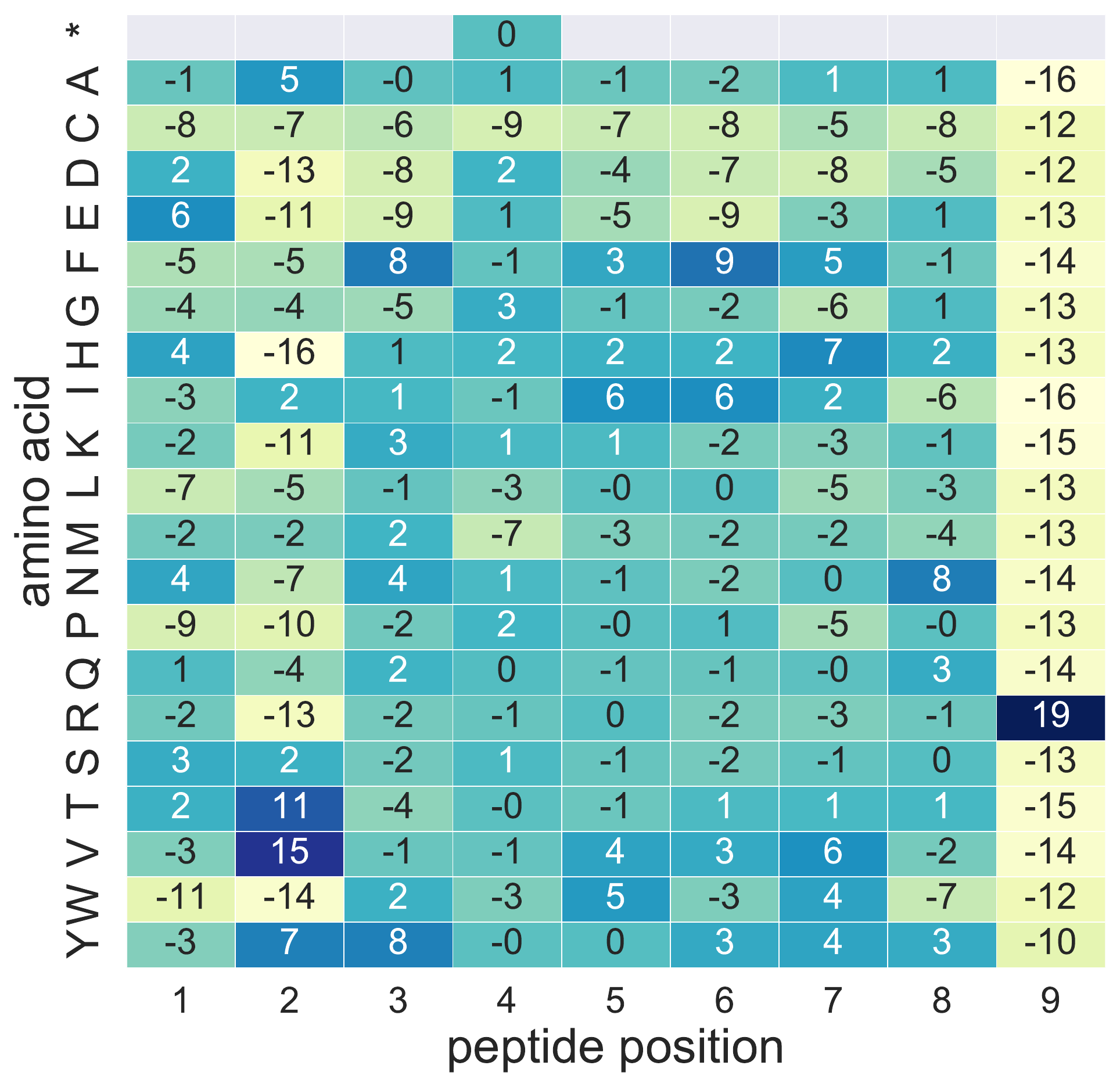}
					\captionof{figure}{Mean \gls{SHAP} values}
					\label{fig:SHAP_\HLAalleleI}
					\end{subfigure}
			        \caption{Motifs and SHAP values for peptide samples presentation on \HLAalleleIname}
				    \label{fig:motif-shap-\HLAalleleIname}
				\end{figure}

				\paragraph{LIME Analysis and 3D structure visualization:} Figure \ref{fig:LIME_\HLAalleleI} shows for each input position, the proportion of examples in which this position had a certain value range of importance ranking (see Subsection \ref{subsubsec:method_Interpretation}). The 9-mer peptide's \glspl{AA} tend to be much more highly ranked than its flanking regions or the AA in the HLA protein. Peptide position 9 is ranked first in roughly half of the examples. The \gls{HLA} pseudo sequence positions display a high variance in their ranking distributions. Some positions tend to be particularly highly ranked - like 63, 73, 77, 97, 116 and 171. \gls{HLA} positions 63, 171 are located in the \gls{HLA}'s A and 77, 116 are located in its F pocket \cite{van_deutekom_zooming_2015} which both are supposed to house a peptide terminus \cite{stryhn_longer_2000}. This might explain what we see in the motifs of Figure \ref{fig:LOGO_\HLAalleleI}. A peptide C-terminus arginine (R) is very common and also the N-terminus shows preference for certain \glspl{AA}.
				
\begin{figure}
				    \centering
				    \begin{subfigure}{0.33\textwidth}
					\includegraphics[width=4cm]{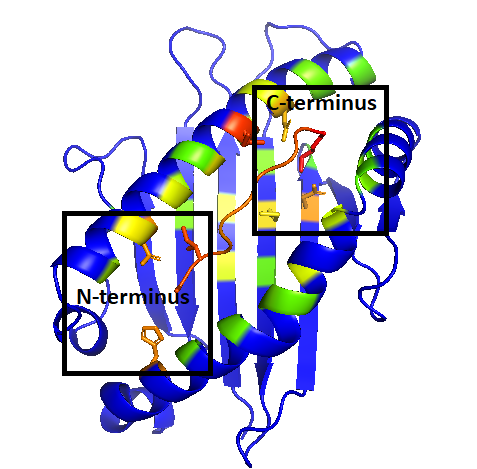}
					\captionof{figure}{Overall perspective}
					\label{fig:3D_\HLAalleleI_total}
					\end{subfigure}
                    \begin{subfigure}{0.32\textwidth}
                    \includegraphics[width=4cm]{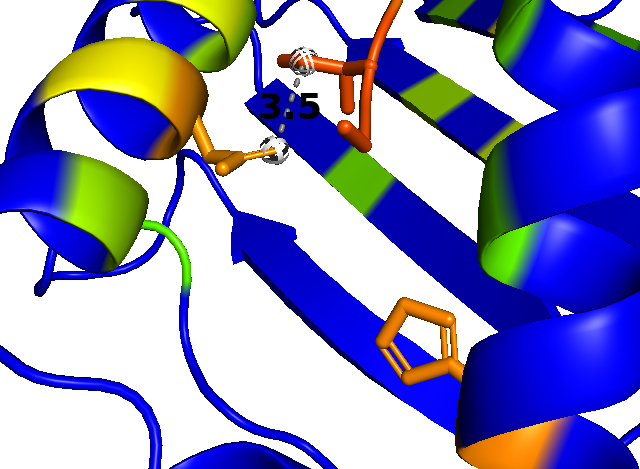}
					\captionof{figure}{Zoom on peptide's N terminus}
					\label{fig:3D_\HLAalleleI_N}
					\end{subfigure}
					\begin{subfigure}{0.32\textwidth}
                    \includegraphics[width=4cm]{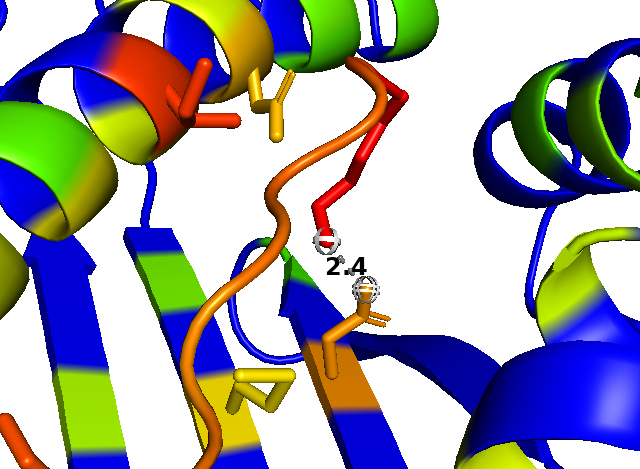}
					\captionof{figure}{Zoom on peptide's C terminus}
					\label{fig:3D_\HLAalleleI_C}
					\end{subfigure}					
			        \caption{PyMOL visualizations of the \HLAalleleIname protein and the peptide}
			        \vspace{-10pt}
				    \label{fig:3D-\HLAalleleI}
				\end{figure}

				The significance of \gls{HLA} positions 73 (in pocket C) and 63 (in pocket B) is less clear. However, in Figure \ref{fig:3D-\HLAalleleI} we display the residues of \gls{HLA} positions 63, 73, 77, 97, 116 and 171 as well as peptide positions 1, 2 and 9. We see, that they can all be found close to the peptide's termini - which aligns with the terminis' high importance values assigned by LIME. We also observe, that the distance between \gls{HLA} position 116 and peptide position 9 is only 2.4 Angstrom in Figure \ref{fig:3D_\HLAalleleI_C}. Whatsmore, position 116 of the MHC protein is a negatively charged aspartate (D), which explains the frequent occurrence of the positively charged arginine (R) at position 9 of presented peptides.  In addition, the distance between \gls{HLA} position 63 and peptide position 2 is only 3.5 Angstrom in Figure \ref{fig:3D_\HLAalleleI_N}. The visualizations clearly show that these high importance features are physically close to each other in 3D space helping us understand how the peptides are presented on MHC-I molecules.
				
	            The peptide flanks are the the least important in Figure \ref{fig:LIME_\HLAalleleI}. This is similar to observations in \cite{odonnell_mhcflurry_2020} where they found that including the peptide flanks resulted in a small but consistent improvement in prediction. Within the flanks, our model attributes most importance to positions closest to the 9-mer peptide. 
				
				\paragraph{Sequence Motifs:} Figure~\ref{fig:LOGO_\HLAalleleI} shows the data and model motifs for \HLAalleleIname. We remind the reader that data motif at the top of Figure~\ref{fig:LOGO_\HLAalleleI} is generated from 100K random 9-mer peptides from the human proteome that are presented (have a presentation score $> 0.5$) with the given HLA allele. Model motif at the bottom of Figure~\ref{fig:LOGO_\HLAalleleI} is generated from 9-mer peptides in our test set that are presented with the given HLA allele.   We find both motifs show a high frequency for $R$ in position 9. \glspl{AA} frequencies in positions 1 and 2 do not match exactly. This might be because they were generated from different background distributions - while the model's motif is based on peptides samples from the whole human genome, the data motif is only based on the proteins actually expressed in the samples of the test \gls{HLA} allele.
				
				\paragraph{SHAP Values:} With regards to \gls{SHAP} values, Figure \ref{fig:SHAP_\HLAalleleI} left side shows the mean \gls{SHAP} value of each \gls{AA} at each peptide position. We see the strong positive mean contribution of R at peptide position 9, making its high frequency in Figures \ref{fig:LOGO_\HLAalleleI} plausible. It also shows high values for V and T at position 2 which are also enriched in the model motif.

	\subsection{\HLAalleleIIname}
				
\begin{figure}
						\includegraphics[width=14.5cm]{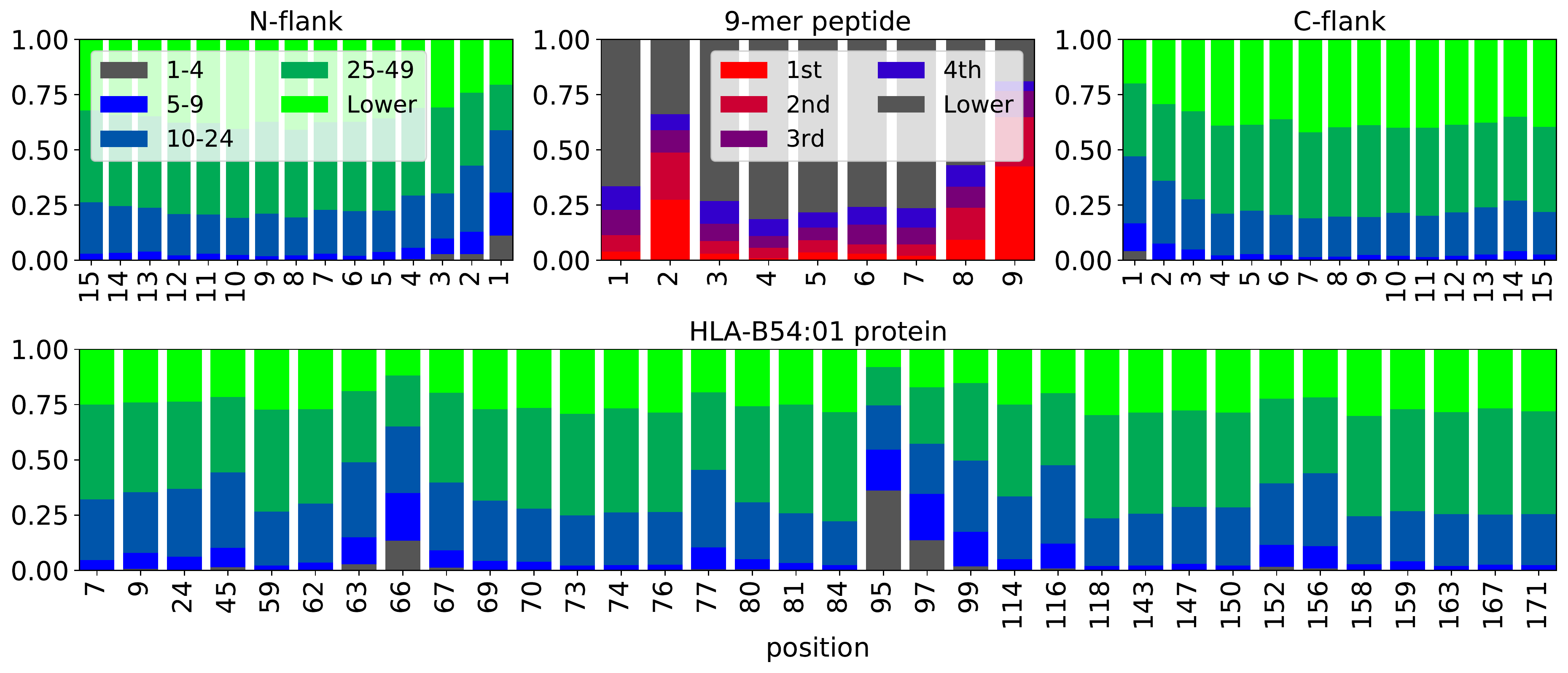} 
						\captionof{figure}{\gls{LIME} feature importance rank distribution for \HLAalleleIIname}
						\label{fig:LIME_\HLAalleleII}
				\end{figure}

\begin{figure}
				    \centering
				    \begin{subfigure}[b]{0.45\textwidth}
					\includegraphics[width=6cm]{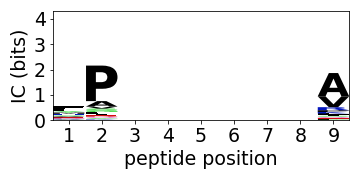}
					\includegraphics[width=6cm]{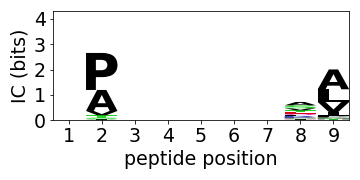} 
					\caption{Data (top) \& model motif}
					\label{fig:LOGO_\HLAalleleII}
					\end{subfigure}
                    \hfill
                    \begin{subfigure}[b]{0.45\textwidth}
					\includegraphics[width=6cm]{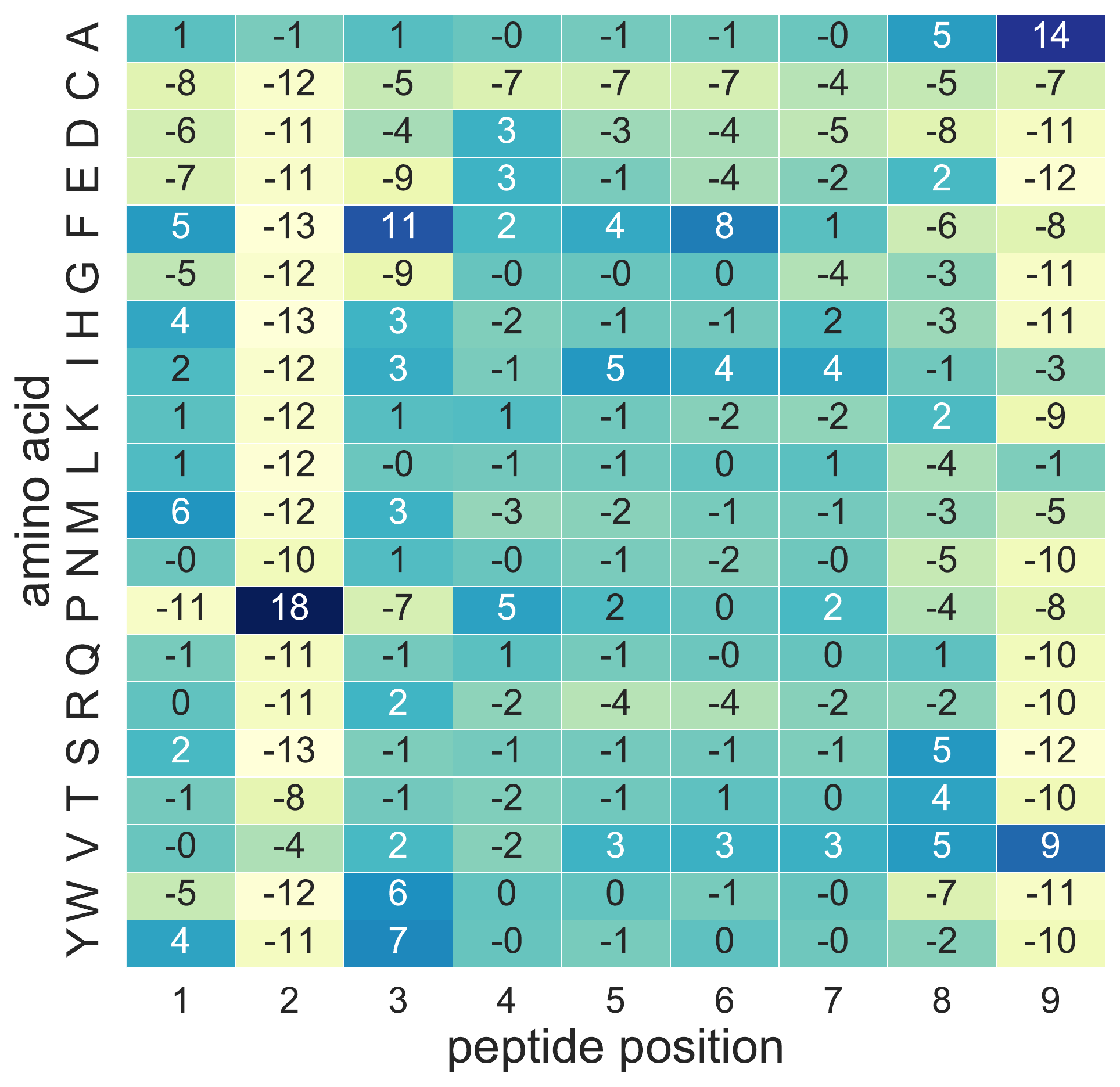}
					\captionof{figure}{Mean \gls{SHAP} values}
					\label{fig:SHAP_\HLAalleleII}
					\end{subfigure}
			        \caption{Motifs and SHAP values for peptide samples presentation on \HLAalleleIIname}
				    \label{fig:motif-shap-\HLAalleleIIname}
				\end{figure}

				\noindent Feature importance ranking generated by \gls{LIME} in Figure \ref{fig:LIME_\HLAalleleII} for peptide presentation with \HLAalleleIIname is similar to that observed in Figure \ref{fig:LIME_\HLAalleleI}. We find again the 9-mer peptide is the most important element followed by the \gls{HLA} allele.  Peptide flanks are the least important. We find two positions in the 9-mer peptide samples to be particularly important for presentation on the \HLAalleleIIname, namely positions 2 and 9. This corresponds to what we see in the motifs in Figure \ref{fig:LOGO_\HLAalleleII} and the \gls{SHAP} values in Figure \ref{fig:SHAP_\HLAalleleII}.
				
\begin{figure}
				    \centering
				    \begin{subfigure}[b]{0.33\textwidth}
					\includegraphics[width=5cm]{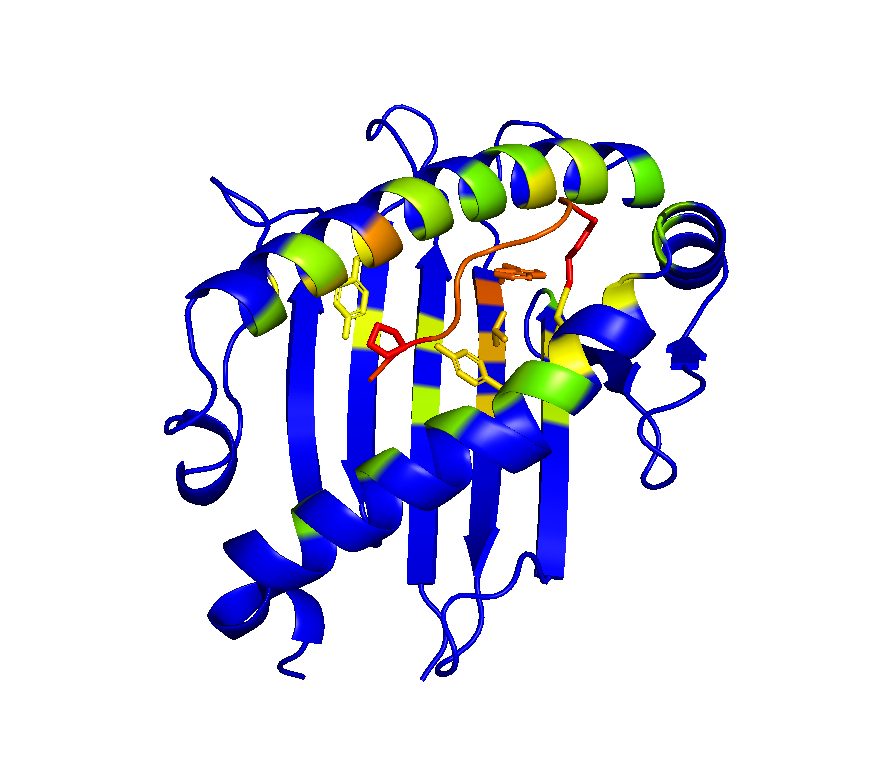}
					\captionof{figure}{Overall perspective}
					\label{fig:3D_\HLAalleleII_total}
					\end{subfigure}
                    \begin{subfigure}[b]{0.31\textwidth}
                    \includegraphics[width=4cm]{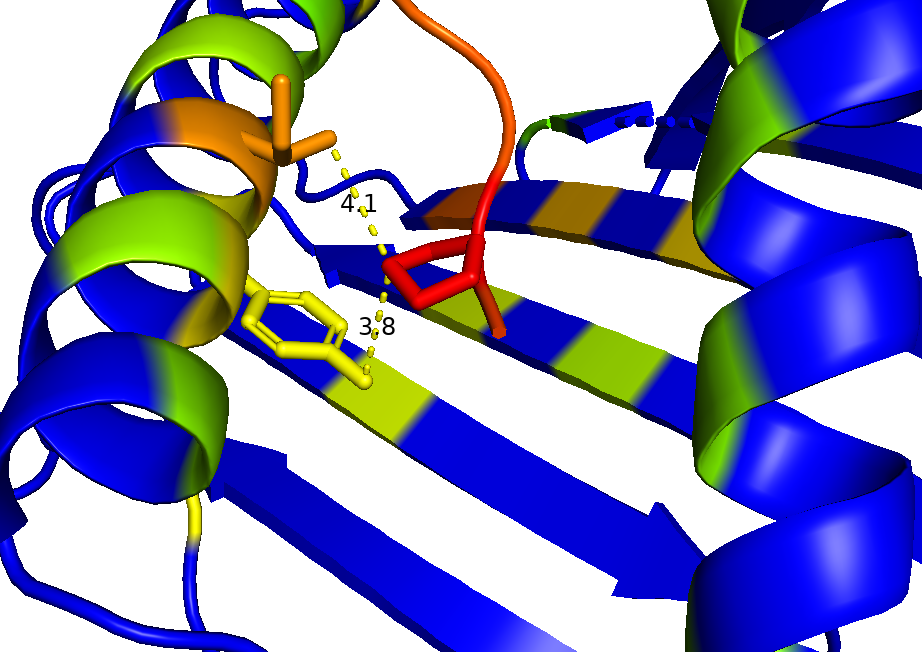}
					\captionof{figure}{Zoom on peptide's N terminus}
					\label{fig:3D_\HLAalleleII_N}
					\end{subfigure}
					\begin{subfigure}[b]{0.31\textwidth}
                    \includegraphics[width=4cm]{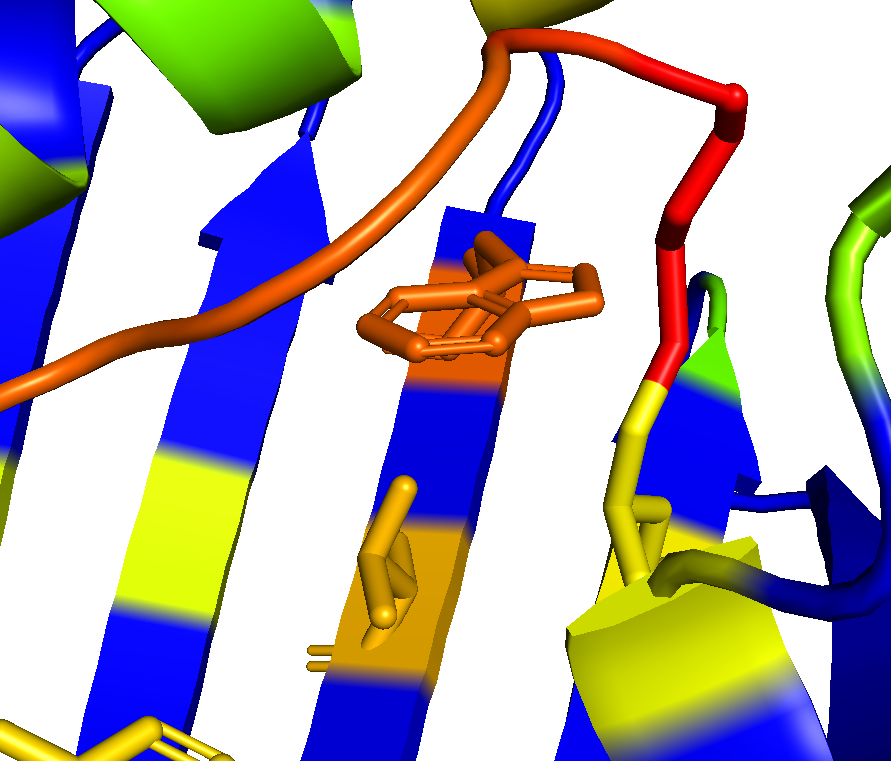}
					\captionof{figure}{Zoom on peptide's C terminus}
					\label{fig:3D_\HLAalleleII_C}
					\end{subfigure}					
					
			        \caption{PyMOL visualizations of the \HLAalleleIIname protein and the peptide}
				    \label{fig:3D-\HLAalleleII}
				\end{figure}

				Figure \ref{fig:LOGO_\HLAalleleII} shows that the peptide position 9 is enriched by alanine, valine and leucine - all of which are hydrophobic. So it is important that also the corresponding positions in the \gls{HLA} protein's F pocket are hydrophobic as well. Figure \ref{fig:3D_\HLAalleleII_N} shows a tryptophan in orange at position 95 and a leucine in yellow at position 116 - both as well hydrophobic. In figure \ref{fig:LIME_\HLAalleleII} we then see that the model indeed gives high importance to those two \gls{HLA} positions.
				
				The \gls{HLA} protein's second most important position is 66 in pocket B. Given the enrichment of peptide position 2 by P and A we speculate that pocket B houses this peptide position. A similar finding was reported in~\cite{liu_major_2011}. Indeed, when we look at figure \ref{fig:3D_\HLAalleleII_N}, we see that the peptide position 2 in red is located between the two \gls{HLA} positions 66 (in orange) and 67 (in yellow). These two are isoleucine and tyrosine - two hydrophobic amino acids.

				\noindent  
				

\section{Conclusion}
\label{section:conclusion}
	We applied interpretability techniques \gls{LIME} and \gls{SHAP} to find that our model learned biologically meaningful importance rankings and feature contributions. We confirmed the interpretations using 3D structure visualizations and sequence motifs of peptides and MHC molecules. We found across HLA alleles, high importance for peptide presentation was given to \glspl{AA} near the N- and C-termini of the peptide and varying \gls{MHC} positions in the A, B and F pockets. In contrast, the peptide flanks showed less importance, which explains why \cite{odonnell_mhcflurry_2020} found that including them only results in a small but consistent model improvement.  The motifs we found using our model followed broadly those observed in the data. As these analysis were all carried out on held out \gls{MHC} proteins, it also demonstrates the generalization ability of our model. 
	
	We have only scratched the surface in attempting to understand peptide presentation with MHC-I molecules by interpreting the results from deep learning models. Visualizations, sequence motifs and interpretability techniques like LIME go hand in hand in helping both computer scientists and biologists understand the application of deep learning models for peptide presentation and the underlying biology. Given the interdisciplinary nature of this field, it is important that future deep learning models that provide accurate predictions in this field are accompanied with explanations and visualisations accessible to biologists and clinicians.

\section{Ethics considerations}
\label{section:ethics}
    We expect our research to contribute to an improved understanding of the \gls{MHC-I} pathway which will enable better customized therapies also for patients with less common \gls{MHC-I} alleles.
    The datasets we worked with did not include any sensitive personally identifiable information. It, however, does not represent the diversity of the global human population. This is exactly the reason, why it is important to develop models that can extrapolate to unseen \gls{MHC} proteins.

\section{Acknowledgments}
The study was supported by the project ‘International Centre for Cancer Vaccine Science’ that is carried out within the International Agendas Programme of the Foundation for Polish Science co-financed by the European Union under the European Regional Development Fund. 

We thank the PL-Grid and CI-TASK Infrastructure, Poland, for providing their hardware and software resources.

\section{Author information}
These authors jointly supervised this work: Ajitha Rajan and Javier Alfaro \\
The co-corresponding authors are: Hans-Christof Gasser, Ajitha Rajan and Javier Alfaro

\printglossaries
\printbibliography

\appendix

\section{Appendix}

	\subsection{Experimental datasources}
    \label{subsec:experimental_datasources}
		As the most restrictive step in epitope presentation is \gls{MHC} binding, measuring the \gls{BA} between a particular \gls{MHC} protein and a peptide \textit{in vitro}, can give us some insight into how likely it is that a particular peptide will be presented to the extracellular environment. Early presentation predictors were, therefore, only trained on this \gls{BA} data. Its biggest disadvantage is, that it is costly to carry out the experiments that then only generate little data \cite{alvarez_nnalign_ma_2019}.

		In the modern high throughput \gls{EL} approach, the whole immunopeptidome of the cell is harvested and then the peptides are identified using mass spectrometry \cite{hunt_characterization_1992}. This identifies many peptides at once. However, it is typically not possible to measure which of the up to six different \gls{MHC} alleles presented which eluted peptide. Monoclonal cell lines, that only express a single \gls{MHC} allele are a potential way around this limitation. There also exist algorithms to identify which allele is responsible for a peptide's presentation. This step is called ``deconvolution'' in the literature. However, it  also often relies on data that can be unambiguously ascribed to a single allele to kickstart it.

		Another disadvantage of the \gls{EL} approach is, that it does not generate definitive negative examples. It only reports about the presence of a peptide at the cell's surface. It cannot assert the absence of a peptide from the individuals immunopeptidome. For example a peptide present in the human genome, might only not be presented by the cell because its protein is not being expressed by the particular cell. Despite it's drawbacks, the high quantity of data generated by this approach will make it the backbone of our examinations.

    \subsection{Data split}
    \label{subsec:data_split}

			\textbf{MHC allele dimension:} As each individual normally has at least one working copy of each \gls{HLA} gene (A/B/C), it is not possible to hold out a full gene. So, we hold out observations on the \gls{HLA} group level (e.g. HLA-A*01, ...). First we count how many observations belong to each \gls{HLA} group (from a donor/cell with at least one \gls{HLA} gene being in the group). We find that the groups are highly unequally represented in the dataset. To find at least 5 groups for each set, we perform the following steps until a satisfactory split is found. 
			
			First, we randomly assign \gls{HLA} groups (and the linked examples) to the validation set until its target number of examples is reached - overriding the standard training set assignment. Then we randomly assign allele groups to the test set until its target number of examples is reached - overriding any earlier assignment. If we assigned too few or too many we repeat. This process ensures that no validation or test \gls{MHC} group enters the training phase and no test \gls{MHC} group enters the validation or training phases.
	
			\textbf{Protein dimension:} Following this, we split off another validation and test set from the remaining training set. This second split is, however, based on an observation's mapped proteins not its linked \gls{MHC} alleles. Due to homology we cannot just split the dataset based on the protein names. There are different approaches to deal with this. We use the python networkx package, which allows to build and explore graph structures. With it we use the Ensembl BioMart paralogue table \cite{ensembl_biomart_2021} to link related genes as well as proteins to their respective gene. We then randomly assign disconnected sub-graphs to the various splits until our target values are reached.
			
            Applying the above, we obtain the partition of our data in Table \ref{tab:split}.
			
			\begin{table}[H]
			\vskip 0mm
			\begin{center}
			\begin{small}
			\begin{sc}
			\begin{tabular}{| l | c | c c | c c |}
			\hline
			Split				& train			& val-prot			& test-prot			& val-mhc			& test-mhc \\
			\hline
			\hline
			Total observations		& \makecell{1,408k \\ (71.8\%)}	 	& \makecell{70k \\ (3.6\%)}		& \makecell{71k \\ (3.6\%)}		& \makecell{204k \\ (10.4\%)}		& \makecell{206k \\ (10.5\%)} \\
			\hline
			Single allele			& \makecell{206k \\ (10.5\%)}		& \makecell{10k \\ (0.5\%)}		& \makecell{11k \\ (0.6\%)}    		& \makecell{24k \\ (1.2\%)}  		& \makecell{43k \\ (2.2\%)} \\
			\hline
			Multiple allele 		& \makecell{1,202k \\ (61.3\%)}		& \makecell{60k \\ (3.1\%)}		& \makecell{60k \\ (3.1\%)} 		& \makecell{181k \\ (9.2\%)}		& \makecell{164k \\ (8.3\%)} \\		

			\hline
			\end{tabular}
			\end{sc}
			\end{small}
			\caption{Observations per dataset split}
			\label{tab:split}
			\end{center}
			\vskip -7mm
			\end{table}

	\subsection{Hyperparameter search and training}
	\label{subsec:hypsearch}
		We searched parameters along 3 dimensions: pooling mechanism, decoys per hit and learning rate. We considered using the $<$cls$>$ tokens embedding, averaging and a classical attention mechanism as pooling mechanism. Further, we compare using datasets enriched by 19 and 99 decoys per observation. Eventually, we tried using 1e-04, 1e-05 and 1e-06 as initial learning rates. For the hyperparameter search, each model was trained on 10\% of the \gls{SA} data for 64,599 steps (one epoch for the 99 decoys per observation datasets and five epochs for the 19 decoys per observation datasets - so both of them have seen the same observations at least once).
		
		We evaluated each model on 10\% of the \gls{MHC}-validation and protein-validation set  \gls{SA} data (using 99 decoys per hit). Appendix Table \ref{tab:hparam_e05} shows the result for the hyper-parameter search using a learning rate of 1e-05 and Appendix Table \ref{tab:hparam_e06} the results for a learning rate of 1e-06. Using a learning rate of 1e-04 would most of the time not result in any detected hits.
		
		Table \ref{tab:hparam_e05} shows various performance metrics (best column values in red) on our two validation sets for the 6 models described above at two points during their training. Different initializations might deliver different results, as mentioned above we didn't use the full dataset and the training will not have converged after 64,559 steps yet. However, given our limited computing resources and the time it takes to train transformers, we had to base our decision on these numbers.  Dependent on the chosen metric, one or the other pooling mechanism and one or the other hits-to-decoys ratio  looks best. In general, the unbalanced dataset (99 decoys per hit) at first (step 12,911) leads to quite poor classifiers in terms of \gls{ROC}-\gls{AUC} and \gls{AP}. However, after a full epoch they are able to make up most of this. In fact, as found by \cite{weiss_learning_2003}, models trained on an unbalanced dataset close to the actual data distribution (99 decoys per observation) show better accuracy. Actually, the models using only 19 decoys per observation have a worse accuracy than just always predicting negative. However, accuracy is not informative and reliable when dealing with highly imbalanced data (e.g. when the majority to minority class ratio is 999:1 a classifier always predicting the majority class  will have 99.9\% accuracy). The models performing best on \gls{ROC}-\gls{AUC} and \gls{AP} all were trained using 19 decoys-per-hit (Appendix Table \ref{tab:hparam_e05}). So we will use this. Using the classification token as input to the head had the best performance (in red) in 5 cases while the attention mechanism only had the best performance in 3 cases and the averaging in 2 cases. We will, therefore, train our final  model using the classification token's output as input to the model's head. Due to time reasons and little improvement we stopped after epoch 5 and use this as our final model (Table \ref{tab:training}).
		
		\begin{table}[H]
			\vskip 3mm
			\begin{center}
			\begin{small}
			\begin{sc}
			\begin{tabular}{| c c | c c c | c c c |}	
			\hline
			\multicolumn{2}{|c|}{after}				& \multicolumn{3}{c|}{val-mhc}		& \multicolumn{3}{c|}{val-protein} \\
			epochs			& steps			& \gls{AP}	& {\tiny\makecell{\gls{ROC}\\ \gls{AUC}}}	& Acc°  	& \gls{AP} 	& {\tiny\makecell{\gls{ROC}\\\gls{AUC}}}	& Acc° \\
			\hline
			\hline
			1 & 128494 & 0.571 & 0.966 & 0.989 & 0.667 & 0.985 & 0.988 \\
			\hline
			2 & 1008417 & 0.646 & 0.976 & 0.992 & 0.762 & 0.993 & 0.993 \\
			\hline
			3 & 1888340 & 0.671 & 0.978 & 0.993 & 0.767 & 0.993 & 0.994 \\
			\hline
			4 & 2768263 & 0.673 & 0.978 & 0.992 & 0.768 & 0.993 & 0.993 \\
			\hline
			5 & 3648186 & 0.683 & 0.979 & 0.992 & 0.765 & 0.993 & 0.994 \\			
			\hline
			\end{tabular}
			\end{sc}
			\end{small}
			\caption{Performance comparison during training (°... accuracy)}
			\label{tab:training}
			\end{center}
			\vskip -7mm
		\end{table}	

		\begin{table}[H]
			\vskip 3mm
			\begin{center}
			\begin{small}
			\begin{sc}
			\begin{tabular}{| c | c | c | c c c | c c c |}
			\hline
			\multirow{2}{*}{Pooling}		& \multirow{2}{*}{Decoys}		& after				& \multicolumn{3}{c|}{val-mhc}		& \multicolumn{3}{c|}{val-protein} \\
								&						& x steps				& {\tiny\makecell{\gls{ROC}\\ \gls{AUC}}}	& \gls{AP}		& Acc°  	& {\tiny\makecell{\gls{ROC}\\ \gls{AUC}}}	& \gls{AP}		& Acc° \\
			\hline
			\hline
			\multirow{4}{*}{Cls}		& \multirow{2}{*}{19}			& 12911				& 0.938 & 0.319 & 0.986 & 0.956 & 0.447 & 0.987 \\	
								& 						& 64559				& {\color{red}0.949} & 0.394 & 0.982 & {\color{red}0.964} & {\color{red}0.552} & 0.983 \\
								\cline{2-9}
								& \multirow{2}{*}{99}			& 12911				& 0.823 & 0.062 & 0.990 & 0.917 & 0.232 & 0.990 \\
								& 						& 64559				& 0.945 & 0.349 & {\color{red}0.991} & 0.960 & 0.492 & {\color{red}0.992} \\
			\cline{1-9}
			\multirow{4}{*}{Attn}		& \multirow{2}{*}{19}			& 12911				& 0.936 & 0.298 & 0.973 & 0.957 & 0.447 & 0.982 \\
								& 						& 64559				& 0.941 & {\color{red}0.410} & 0.983 & {\color{red}0.964} & 0.535 & 0.982 \\
								\cline{2-9}
								& \multirow{2}{*}{99}			& 12911				& 0.792 & 0.044 & 0.990 & 0.884 & 0.184 & 0.990 \\
								& 						& 64559				& 0.938 & 0.315 & 0.990 & 0.958 & 0.496 & {\color{red}0.992} \\
			\cline{1-9}
			\multirow{4}{*}{Avg}		& \multirow{2}{*}{19}			& 12911				& 0.926 & 0.252 & 0.982 & 0.954 & 0.417 & 0.985 \\
								& 						& 64559				& {\color{red}0.949} & 0.369 & 0.977 & 0.960 & 0.506 & 0.980 \\
								\cline{2-9}
								& \multirow{2}{*}{99}			& 12911				& 0.689 & 0.026 & 0.990 & 0.835 & 0.145 & 0.990 \\
								& 						& 64559				& 0.924 & 0.287 & 0.990 & 0.956 & 0.471 & {\color{red}0.992} \\
	
			\hline
			\end{tabular}
			\end{sc}
			\end{small}
			\caption{Performance comparison for a learning rate of 1e-05 (°... accuracy)}
			\label{tab:hparam_e05}
			\end{center}
			\vskip -7mm
		\end{table}	
	
		\begin{table}[H]
			\vskip 3mm
			\begin{center}
			\begin{small}
			\begin{sc}
			\begin{tabular}{| c | c | c | c c c | c c c |}
			\hline
			\multirow{2}{*}{Pooling}		& \multirow{2}{*}{Decoys}		& after				& \multicolumn{3}{c|}{val-mhc}		& \multicolumn{3}{c|}{val-protein} \\
								&						& x steps				& {\tiny\makecell{ROC\\AUC}}	& \gls{AP}	& Acc°  	& {\tiny\makecell{ROC\\AUC}}	& \gls{AP}	& Acc° \\
			\hline
			\hline
			\multirow{4}{*}{Cls}		& \multirow{2}{*}{19}			& 12911				& 0.660 & 0.020 & 0.990 & 0.771 & 0.061 & 0.990 \\	
								& 						& 64559				& 0.902 & 0.209 & 0.974 & 0.943 & 0.356 & 0.975 \\
								\cline{2-9}
								& \multirow{2}{*}{99}			& 12911				& 0.584 & 0.013 & 0.990 & 0.689 & 0.033 & 0.990 \\
								& 						& 64559				& 0.721 & 0.027 & 0.990 & 0.835 & 0.119 & 0.990 \\
			\cline{1-9}
			\multirow{4}{*}{Attn}		& \multirow{2}{*}{19}			& 12911				& 0.649 & 0.018 & 0.990 & 0.774 & 0.073 & 0.990 \\
								& 						& 64559				& 0.902 & 0.210 & 0.975 & 0.937 & 0.349 & 0.975 \\
								\cline{2-9}
								& \multirow{2}{*}{99}			& 12911				& 0.594 & 0.014 & 0.990 & 0.694 & 0.034 & 0.990 \\
								& 						& 64559				& 0.731 & 0.029 & 0.990 & 0.829 & 0.132 & 0.990 \\
			\cline{1-9}
			\multirow{4}{*}{Avg}		& \multirow{2}{*}{19}			& 12911				& 0.657 & 0.020 & 0.990 & 0.776 & 0.081 & 0.990\\
								& 						& 64559				& 0.892 & 0.202 & 0.969 & 0.937 & 0.349 & 0.970 \\
								\cline{2-9}
								& \multirow{2}{*}{99}			& 12911				& 0.601 & 0.016 & 0.990 & 0.700 & 0.037 & 0.990 \\
								& 						& 64559				& 0.733 & 0.031 & 0.990 & 0.840 & 0.134 & 0.990 \\
	
			\hline
			\end{tabular}
			\end{sc}
			\end{small}
			\caption{Performance comparison for a learning rate of 1e-06 (°... accuracy)}
			\label{tab:hparam_e06}
			\end{center}
			\vskip -7mm
		\end{table}

	\subsection{Evaluation and Benchmarking}
		To shed light on what our model has learnt and to assess its quality, we performed:\\
		\textbf{- Evaluation} on the test sets\\
		\textbf{- Comparison} of our model to MHCflurry and NetMHCpan\\
		\textbf{- Model interpretation} by \gls{LIME} analysis of peptide, flanks and pseudo sequence feature importances, using motifs and by \gls{SHAP} analysis of peptide \glspl{AA} contributions (see)
		
		\subsection{Evaluation on the test set}
	    Table \ref{tab:test} shows the test set performance of our selected final model (epoch 5 in Table \ref{tab:training}). We see that the values are not very different.

		\begin{table}[H]
			\vskip 3mm
			\begin{center}
			\begin{small}
			\begin{sc}
			\begin{tabular}{| c c | c c c | c c c |}	
			\hline
			\multicolumn{2}{|c|}{after}				& \multicolumn{3}{c|}{test-mhc}		& \multicolumn{3}{c|}{test-protein} \\
			epochs			& steps			& \gls{AP}		& {\tiny\makecell{\gls{ROC}\\ \gls{AUC}}}	& Acc°  	& \gls{AP} 	& {\tiny\makecell{\gls{ROC}\\ \gls{AUC}}}	&  Acc° \\
			\hline
			\hline
			\hline
			5 & 3648186 & 0.704 & 0.981 & 0.993 & 0.755 & 0.992 & 0.993 \\			
			\hline
			\end{tabular}
			\end{sc}
			\end{small}
			\caption{Performance on the test set (°... accuracy)}
			\label{tab:test}
			\end{center}
			\vskip -7mm
		\end{table}

	\subsection{Comparison to MHCflurry and NetMHCpan}
		\label{subsec:method_comparison}
		The MULTIALLELIC benchmark dataset of MHCflurry consists of 9,158,100 examples. Each has a peptide, N-flank (15 \gls{AA}), C-flank (15 \gls{AA}), up to six \gls{HLA} alleles as well as the predictions of NetMHCpan, MixMHCpred and MHCflurry for the example. \cite{odonnell_mhcflurry_2020} generated this dataset from 11 studies using \gls{EL} data. For each hit they randomly generated 99 decoys. A more detailed description and the full dataset is available in \cite[Supplement Data S1]{odonnell_mhcflurry_2020}. We run two evaluations on this - one on the whole dataset (9,158,100 examples) and one for which we removed examples of peptides that were already part of our training dataset (2,781,898 examples). For these we predict our model's presentation score, calculate performance metrics and plot \gls{PR} curves for MHCflurry, NetMHCpan and our model (ImmunoBERT).

	\subsection{Benchmarking}
	\label{subsec:results_benchmarking}

		Unluckily, there are no generally agreed upon standard benchmarking datasets available in our domain. However, \cite{odonnell_mhcflurry_2020} have curated a benchmark dataset. We ran the below analysis on the full dataset as well as on one in which we exclude peptides from our training set. We calculated the \gls{AP} as well as the \gls{ROC}-\gls{AUC} for MHCflurry, NetMHCpan and ImmunoBERT. We also plotted the \gls{PR}-curves for them (Table \ref{tab:benchmark_metrics}). The models were trained on different datasets. So any judgement about the advantageousness of the architectures is not valid. However, the comparison is useful to compare the practical predictive power of the models.

		For both datasets, our model shows an \gls{AP} in between MHCflurry and NetMHCpan. In particular, it does well for thresholds corresponding to intermediate recall levels. However, it achieves less \gls{ROC}-\gls{AUC} than the others, possibly caused by the sharp drop in performance for higher recall values. The performance on the reduced set is far worse for all models. So, also the other two models might have already seen similar peptides as were removed during their training. As we explicitly only removed ours, this skews the reduced dataset against our model.
		
		We decided to generate decoys once and show the model the same decoys in each epoch. This was done to ensure reproducibility of results. In contrast, NetMHCpan and MHCflurry resample decoys each epoch \cite{odonnell_mhcflurry_2020}. In hindsight, this might be a better design choice and might have led to later convergence during training of our model.

		\begin{table}[H]
			\vskip 3mm
			\begin{center}
			\begin{small}
			\begin{sc}
			\begin{tabular}{| c | l | c c | c | c c |}
			\hline
			& Model 	/ Metric			& \gls{AP}		& {\tiny\makecell{\gls{ROC}\\\gls{AUC}}} & \gls{PR}-Curve 	 \\
			\hline
			\hline
			\multirow{6}{*}{\rotatebox{90}{Full Dataset}}
			& NetMHCpan			& 0.327		& 0.916 & \multirow{3}{*}{\includegraphics[width=4.0cm]{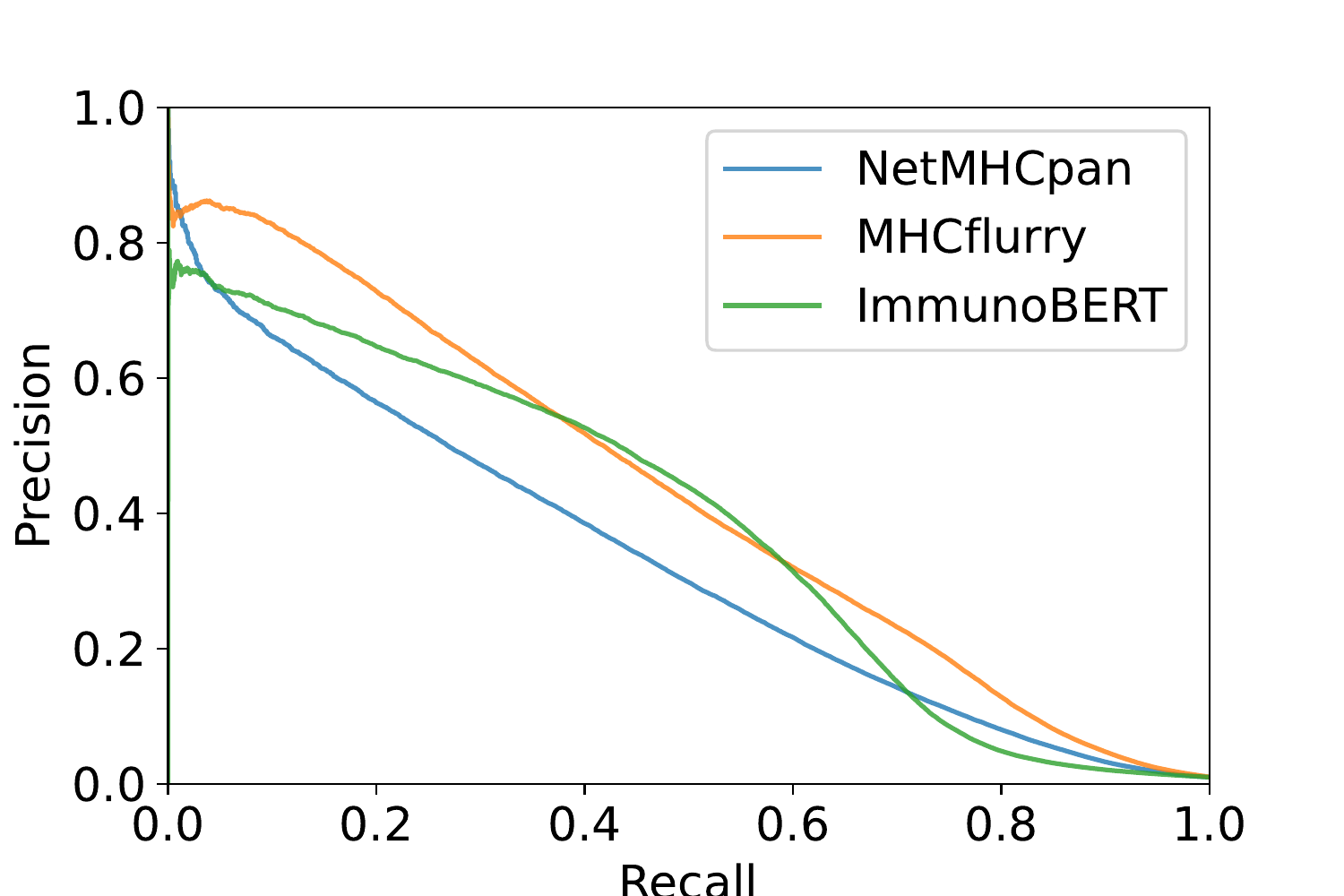} } 	\\	
			& MHCflurry				& 0.427		& 0.938 & 											\\	
			& ImmunoBERT			& 0.383		& 0.893 & 										\\
			& 					&			&	  & 											\\
			& 					&			&	  & 											\\
			& 					&			&	  & 											\\
			& 					&			&	  & 											\\
			& 					&			&	  & 											\\
			\hline
			\multirow{6}{*}{\rotatebox{90}{Reduced Dataset}}
			& NetMHCpan			& 0.151		& 0.873 & \multirow{3}{*}{\includegraphics[width=4.0cm]{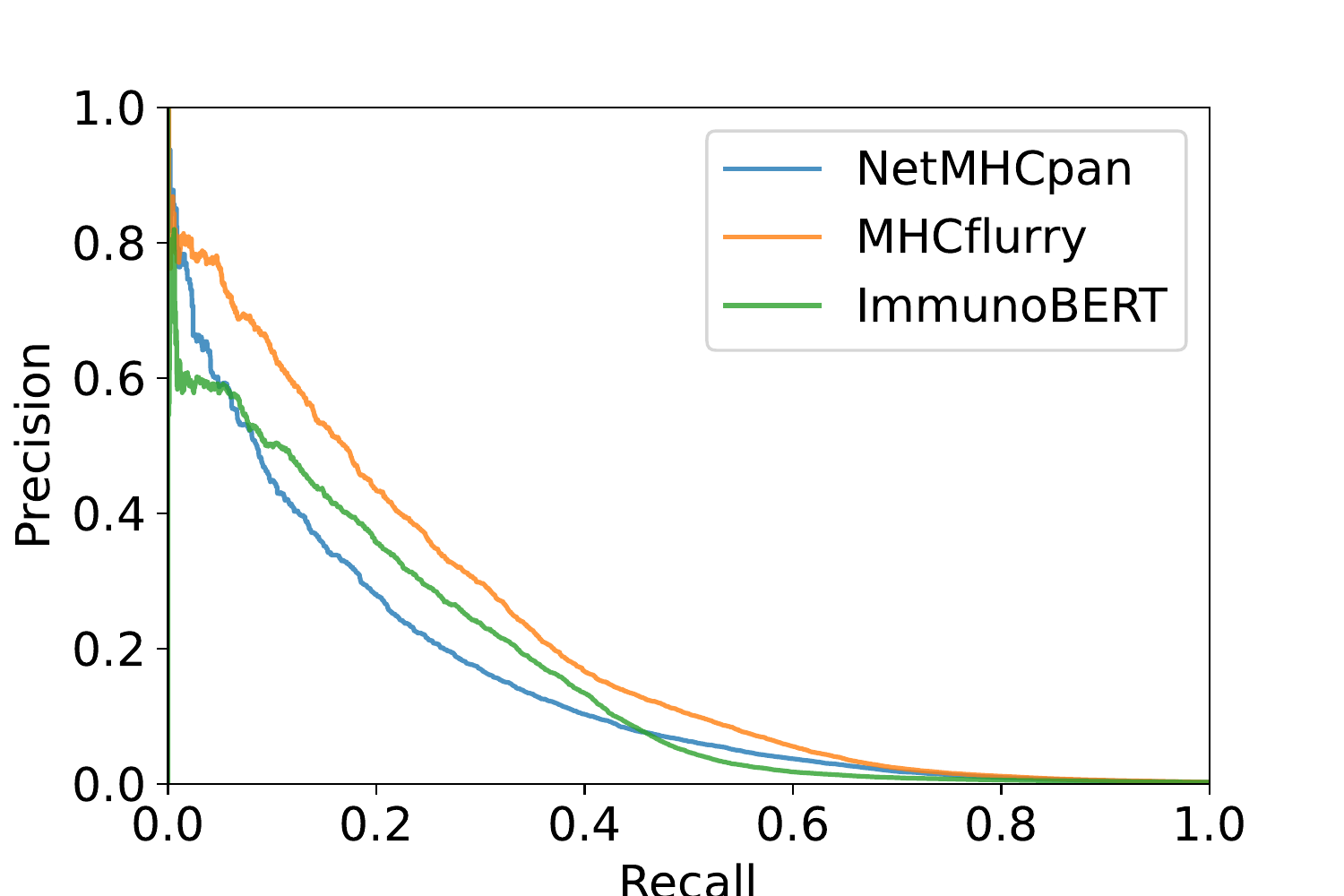} } \\	
			& MHCflurry				& 0.215		& 0.890 & 											\\	
			& ImmunoBERT			& 0.163		& 0.831 & 										\\
			& 					&			&	  & 											\\
			& 					&			&	  & 											\\
			& 					&			&	  & 											\\
			& 					&			&	  & 											\\
			& 					&			&	  & 											\\
			\hline
			\end{tabular}
			\end{sc}
			\end{small}
			\caption{Performance comparison on benchmark dataset}
			\label{tab:benchmark_metrics}
			\end{center}
			\vskip -7mm
		\end{table}	

	\subsection{Interpretation}
		On the following pages, the charts for the remaining \gls{HLA} alleles that were note selected for detailed discussion can be found.

		\newpage
		\renewcommand{\HLAallele}{HLA-A3303}
		\renewcommand{\HLAalleleName}{HLA-A*33:03}
		\subsection{\HLAalleleName}
				\noindent \begin{minipage}[c][4.2cm][t]{7cm}
					\begin{center}
					\includegraphics[width=4cm]{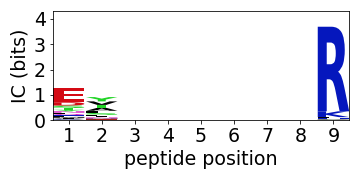}
					\includegraphics[width=4cm]{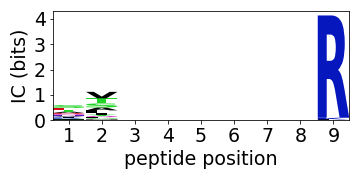} 
					\end{center}
				\end{minipage}
				\begin{minipage}[c][4.2cm][t]{7cm}
					\begin{center}
					\includegraphics[width=4.5cm]{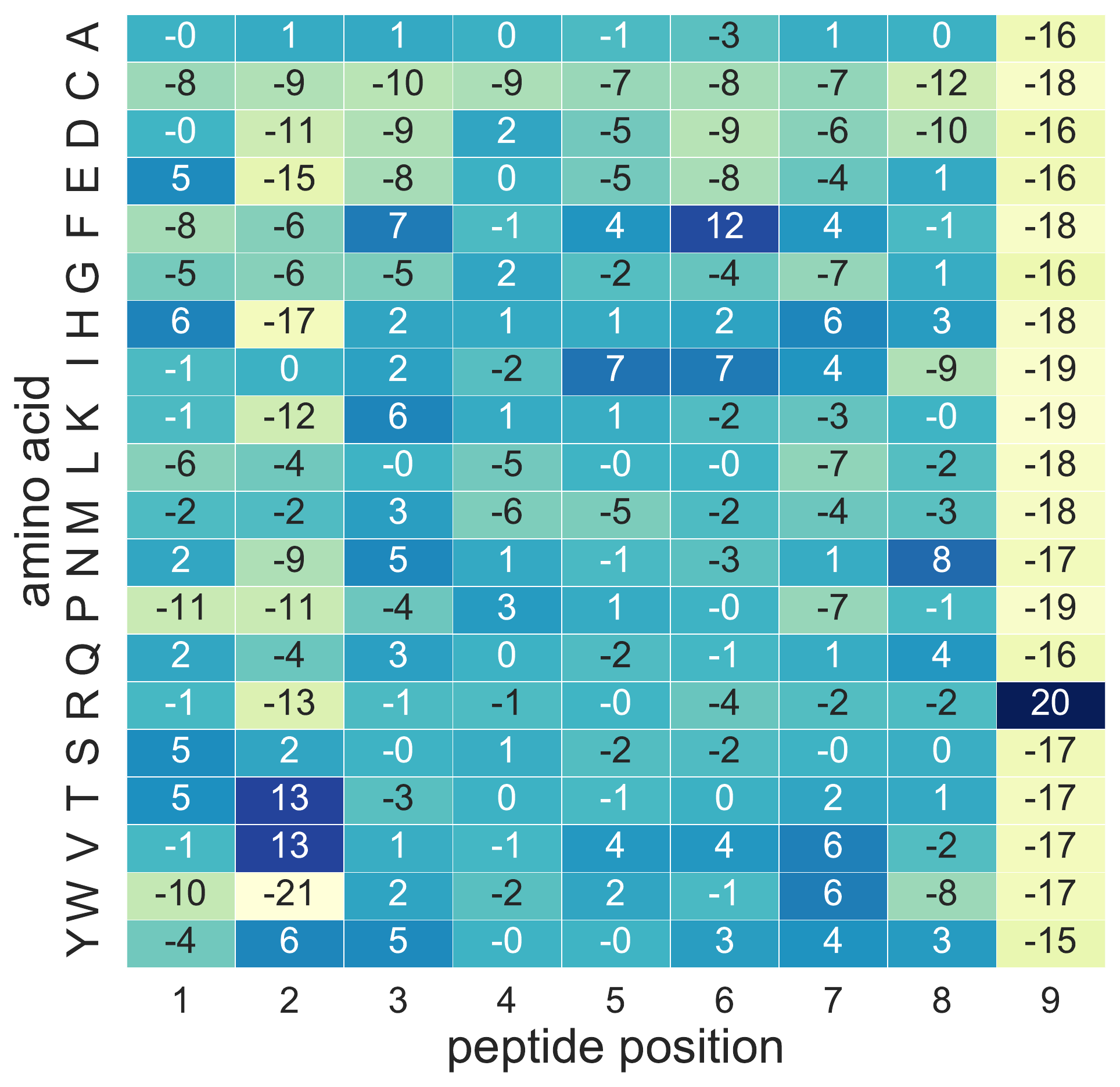}
					\end{center}
				\end{minipage}
				
				\noindent \begin{figure}[H] {\centering
						\includegraphics[width=11cm]{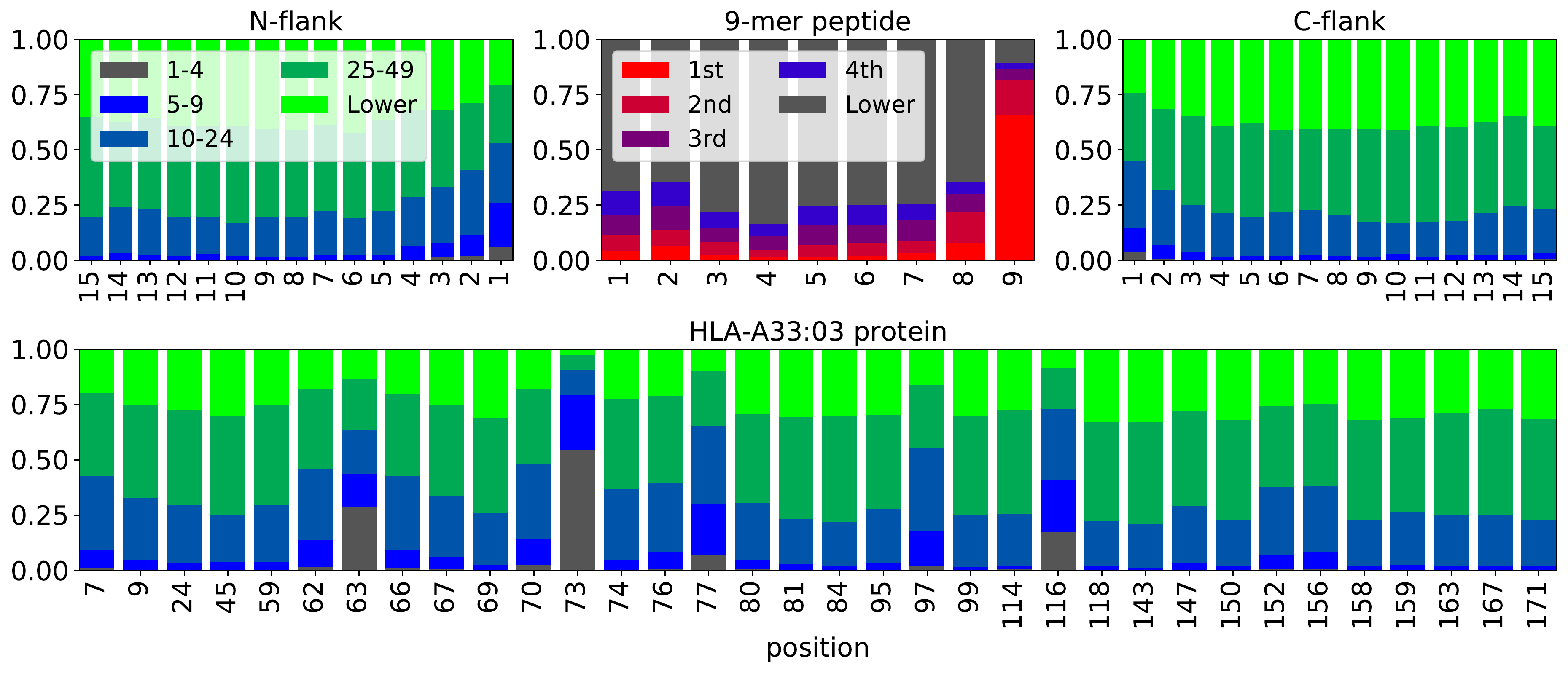} 
						\captionof{figure}{Motifs (top left), mean \gls{SHAP} values (top right) and \gls{LIME} feature ranks}
						\label{fig:LIME_\HLAallele} }
				\end{figure}

		\renewcommand{\HLAallele}{HLA-A3601}
		\renewcommand{\HLAalleleName}{HLA-A*36:01}
		\subsection{\HLAalleleName}
				\noindent \begin{minipage}[c][4.2cm][t]{7cm}
					\begin{center}
					\includegraphics[width=4cm]{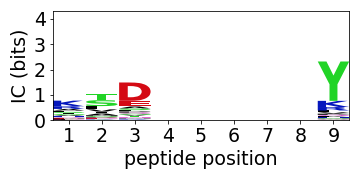}
					\includegraphics[width=4cm]{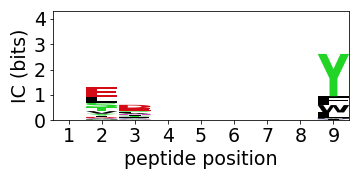} 
					\end{center}
				\end{minipage}
				\begin{minipage}[c][4.2cm][t]{7cm}
					\begin{center}
					\includegraphics[width=4.5cm]{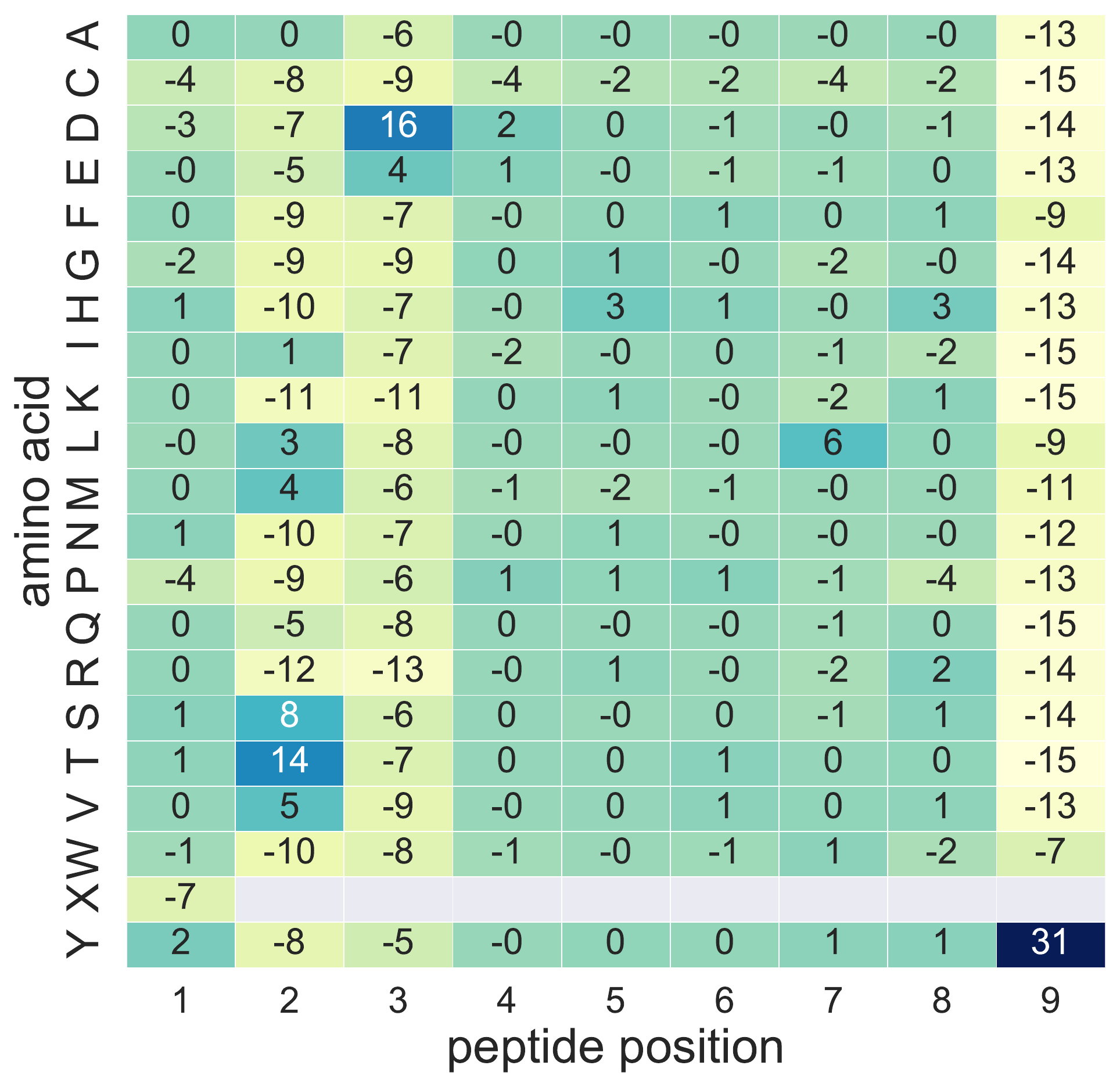}
					\end{center}
				\end{minipage}
				
				\noindent \begin{figure}[H] {\centering
						\includegraphics[width=11cm]{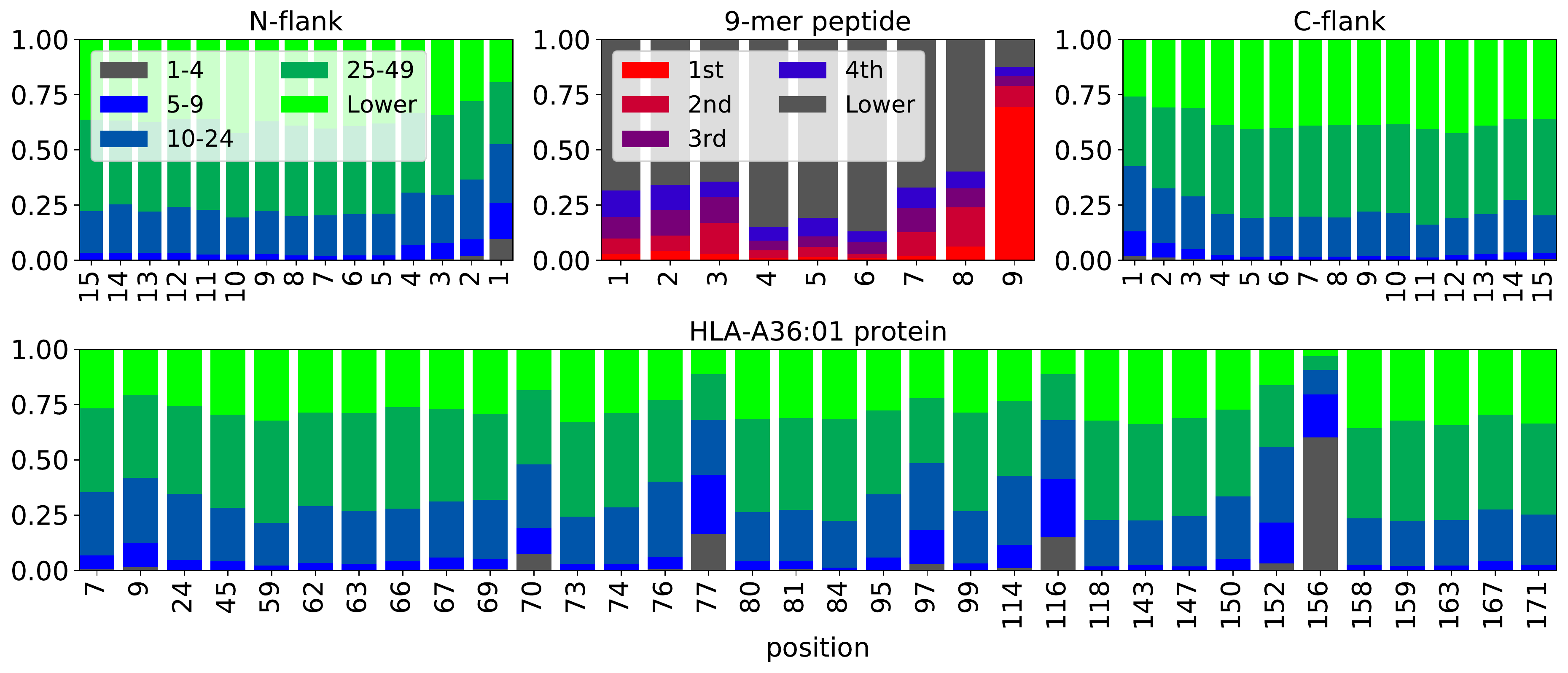} 
						\captionof{figure}{Motifs (top left), mean \gls{SHAP} values (top right) and \gls{LIME} feature ranks}
						\label{fig:LIME_\HLAallele} }
				\end{figure}

		\renewcommand{\HLAallele}{HLA-A7401}
		\renewcommand{\HLAalleleName}{HLA-A*74:01}
		\subsection{\HLAalleleName}
				\noindent \begin{minipage}[c][4.2cm][t]{7cm}
					\begin{center}
					\includegraphics[width=4cm]{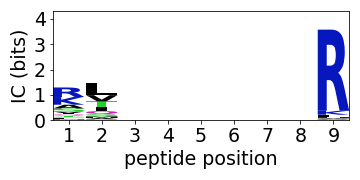}
					\includegraphics[width=4cm]{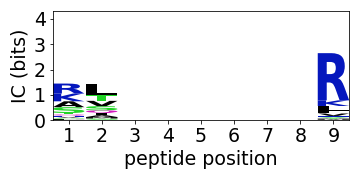} 
					\end{center}
				\end{minipage}
				\begin{minipage}[c][4.2cm][t]{7cm}
					\begin{center}
					\includegraphics[width=4.5cm]{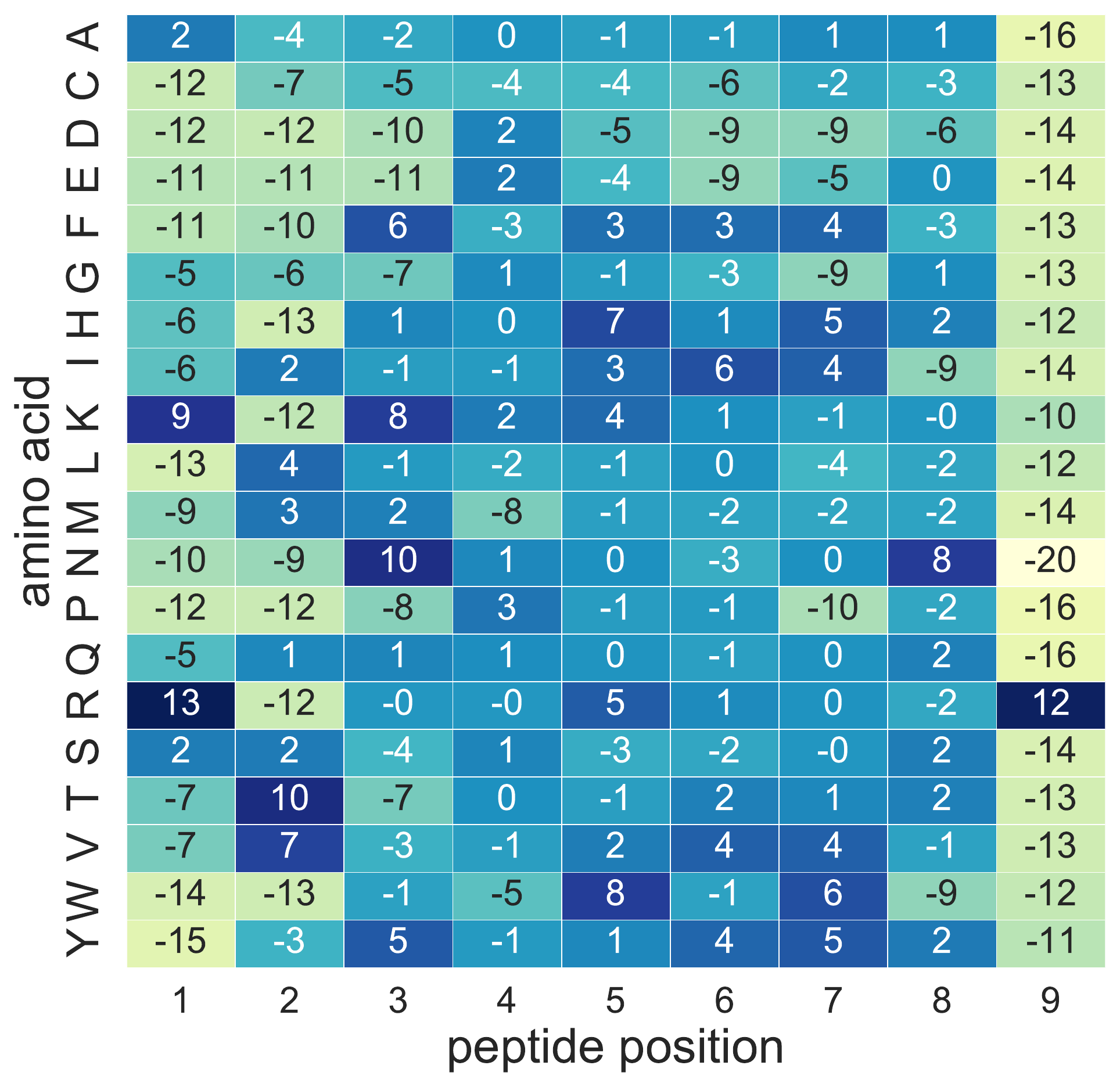}
					\end{center}
				\end{minipage}
				
				\noindent \begin{figure}[H] {\centering
						\includegraphics[width=11cm]{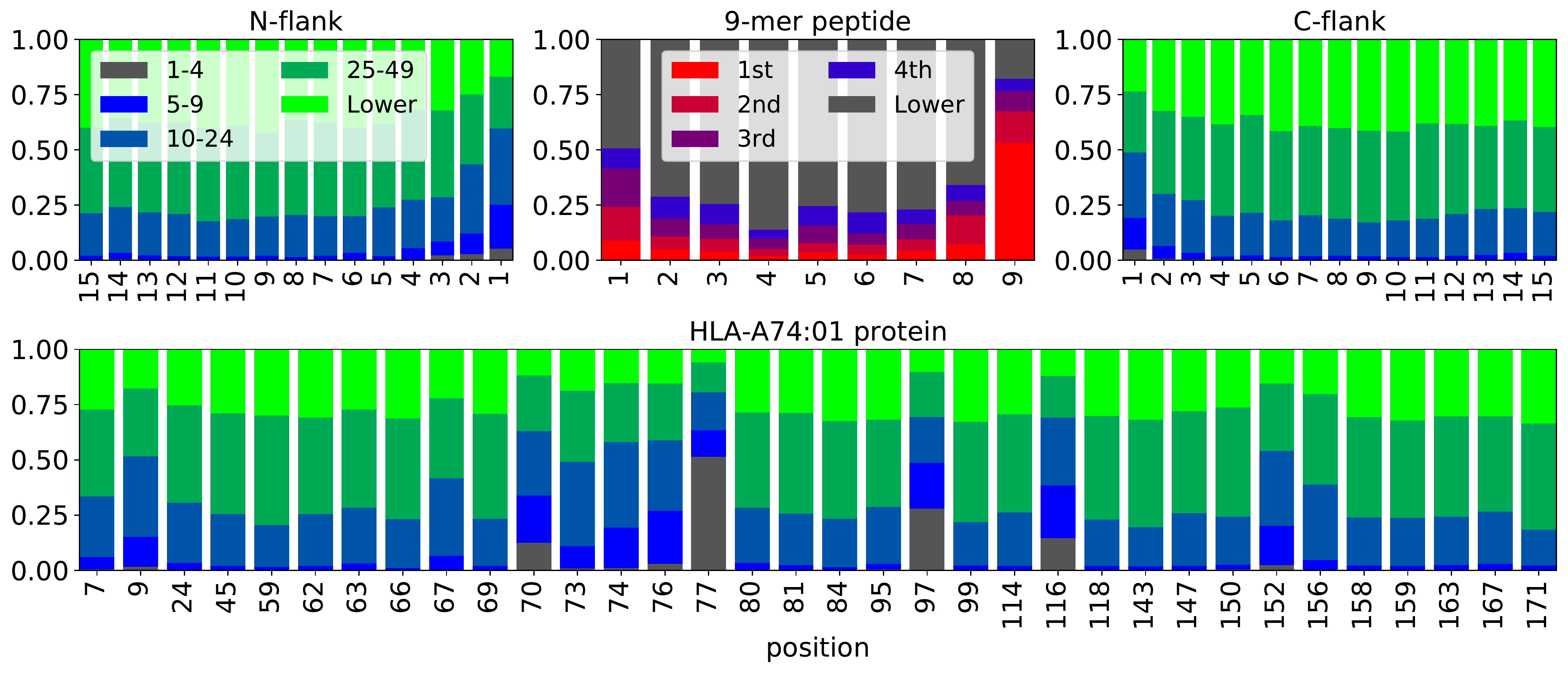} 
						\captionof{figure}{Motifs (top left), mean \gls{SHAP} values (top right) and \gls{LIME} feature ranks}
						\label{fig:LIME_\HLAallele} }
				\end{figure}

		\renewcommand{\HLAallele}{HLA-B3701}
		\renewcommand{\HLAalleleName}{HLA-B*37:01}
		\subsection{\HLAalleleName}
				\noindent \begin{minipage}[c][4.2cm][t]{7cm}
					\begin{center}
					\includegraphics[width=4cm]{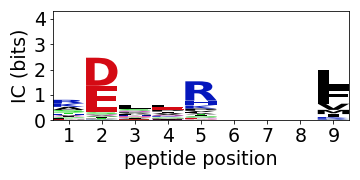}
					\includegraphics[width=4cm]{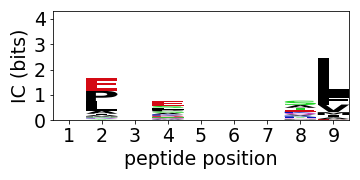} 
					\end{center}
				\end{minipage}
				\begin{minipage}[c][4.2cm][t]{7cm}
					\begin{center}
					\includegraphics[width=4.5cm]{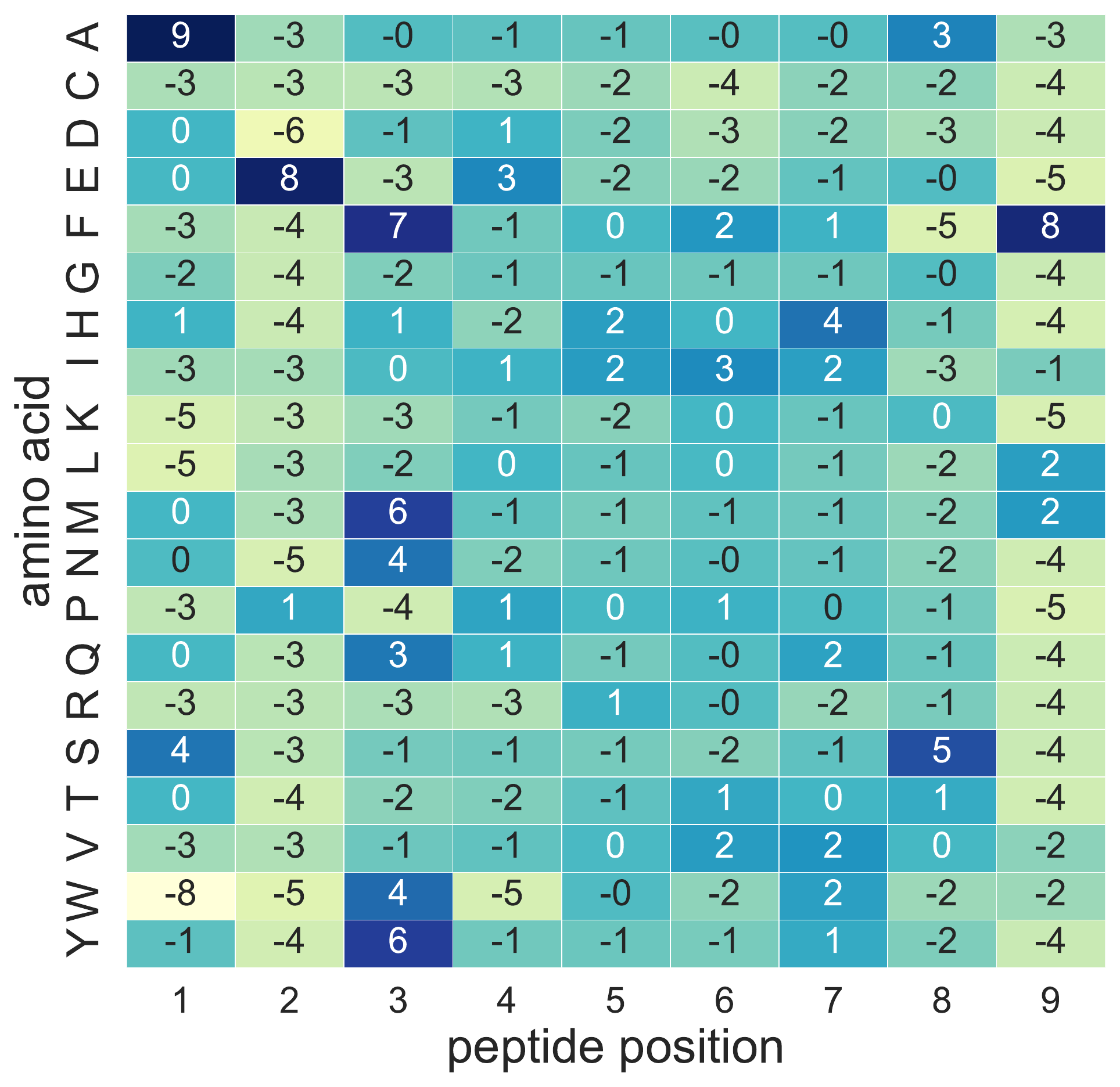}
					\end{center}
				\end{minipage}
				
				\noindent \begin{figure}[H] {\centering
						\includegraphics[width=11cm]{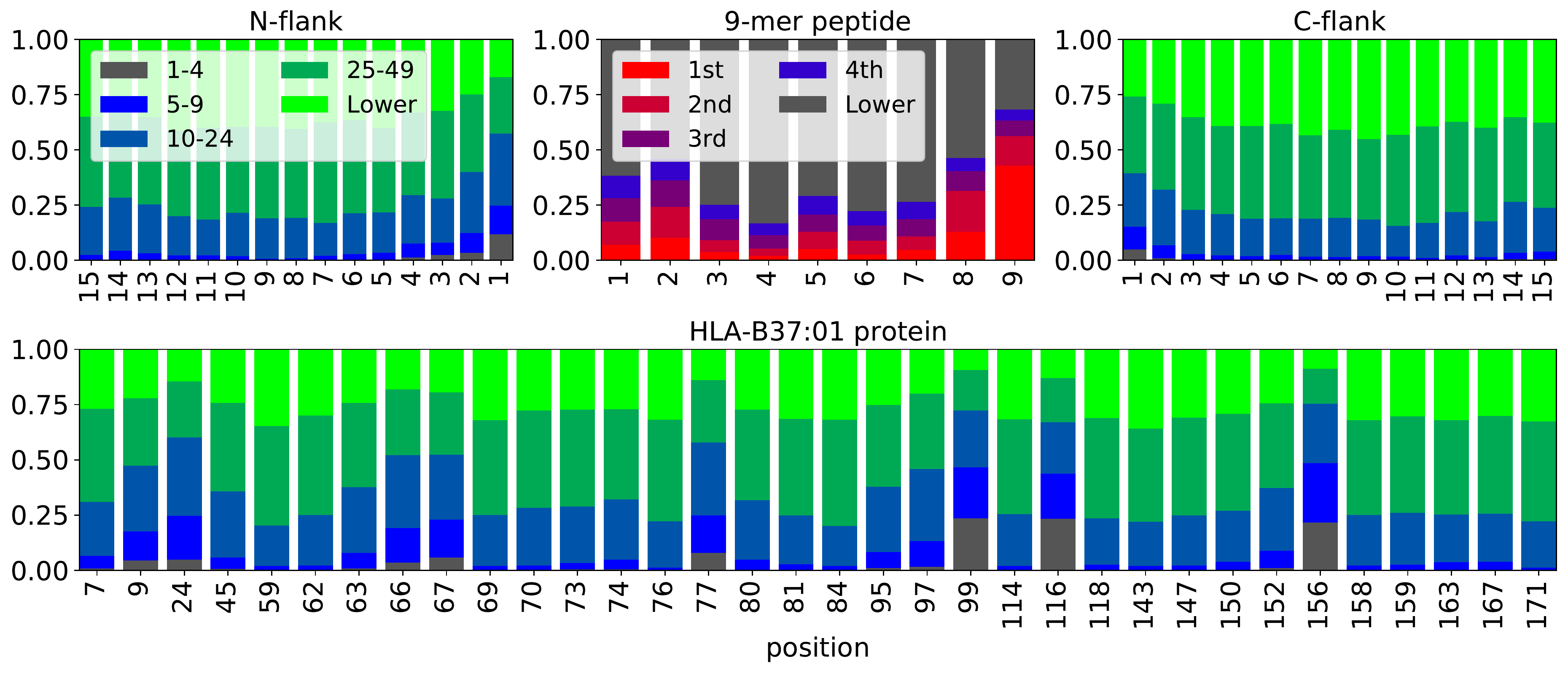} 
						\captionof{figure}{Motifs (top left), mean \gls{SHAP} values (top right) and \gls{LIME} feature ranks}
						\label{fig:LIME_\HLAallele} }
				\end{figure}

		\renewcommand{\HLAallele}{HLA-B4601}
		\renewcommand{\HLAalleleName}{HLA-B*46:01}
		\subsection{\HLAalleleName}
				\noindent \begin{minipage}[c][4.2cm][t]{7cm}
					\begin{center}
					\includegraphics[width=4cm]{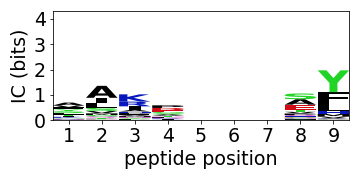}
					\includegraphics[width=4cm]{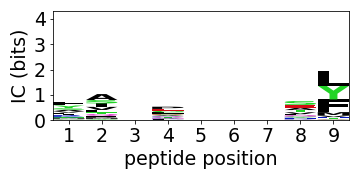} 
					\end{center}
				\end{minipage}
				\begin{minipage}[c][4.2cm][t]{7cm}
					\begin{center}
					\includegraphics[width=4.5cm]{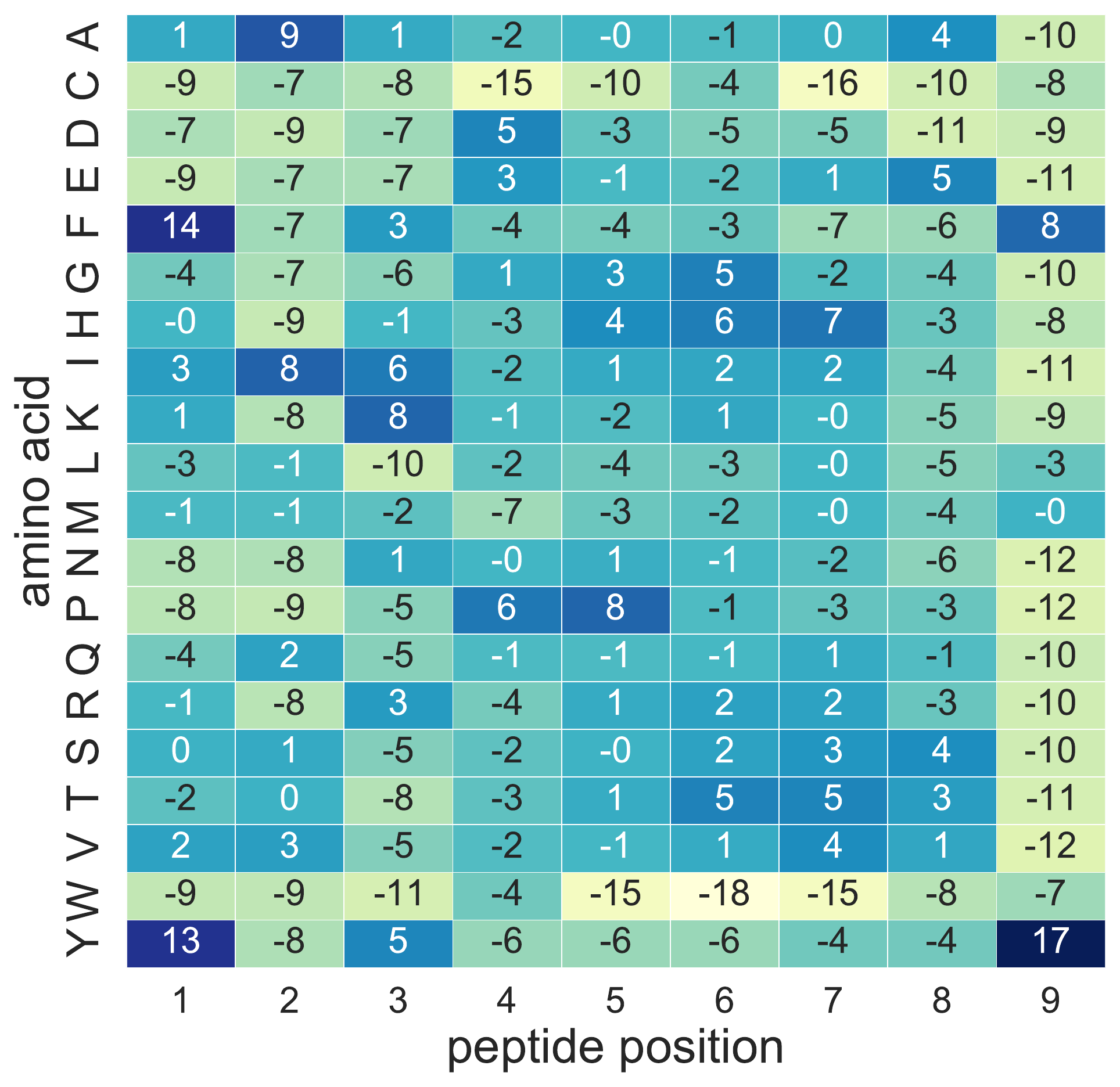}
					\end{center}
				\end{minipage}
				
				\noindent \begin{figure}[H] {\centering
						\includegraphics[width=11cm]{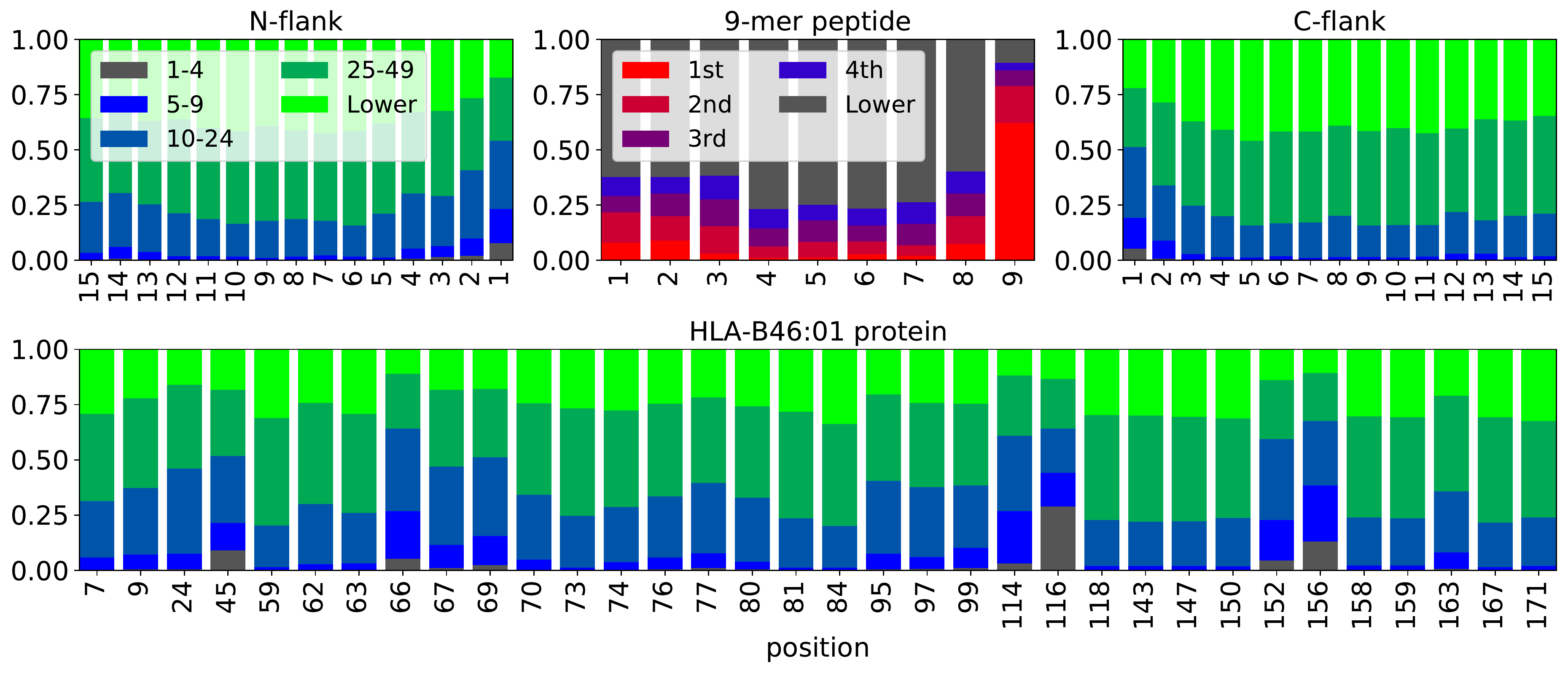} 
						\captionof{figure}{Motifs (top left), mean \gls{SHAP} values (top right) and \gls{LIME} feature ranks}
						\label{fig:LIME_\HLAallele} }
				\end{figure}

		\renewcommand{\HLAallele}{HLA-B5801}
		\renewcommand{\HLAalleleName}{HLA-B*58:01}
		\subsection{\HLAalleleName}
				\noindent \begin{minipage}[c][4.2cm][t]{7cm}
					\begin{center}
					\includegraphics[width=4cm]{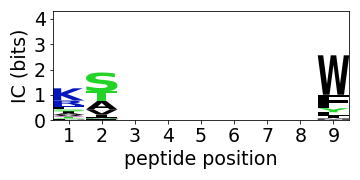}
					\includegraphics[width=4cm]{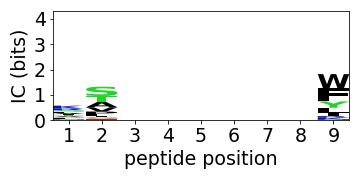} 
					\end{center}
				\end{minipage}
				\begin{minipage}[c][4.2cm][t]{7cm}
					\begin{center}
					\includegraphics[width=4.5cm]{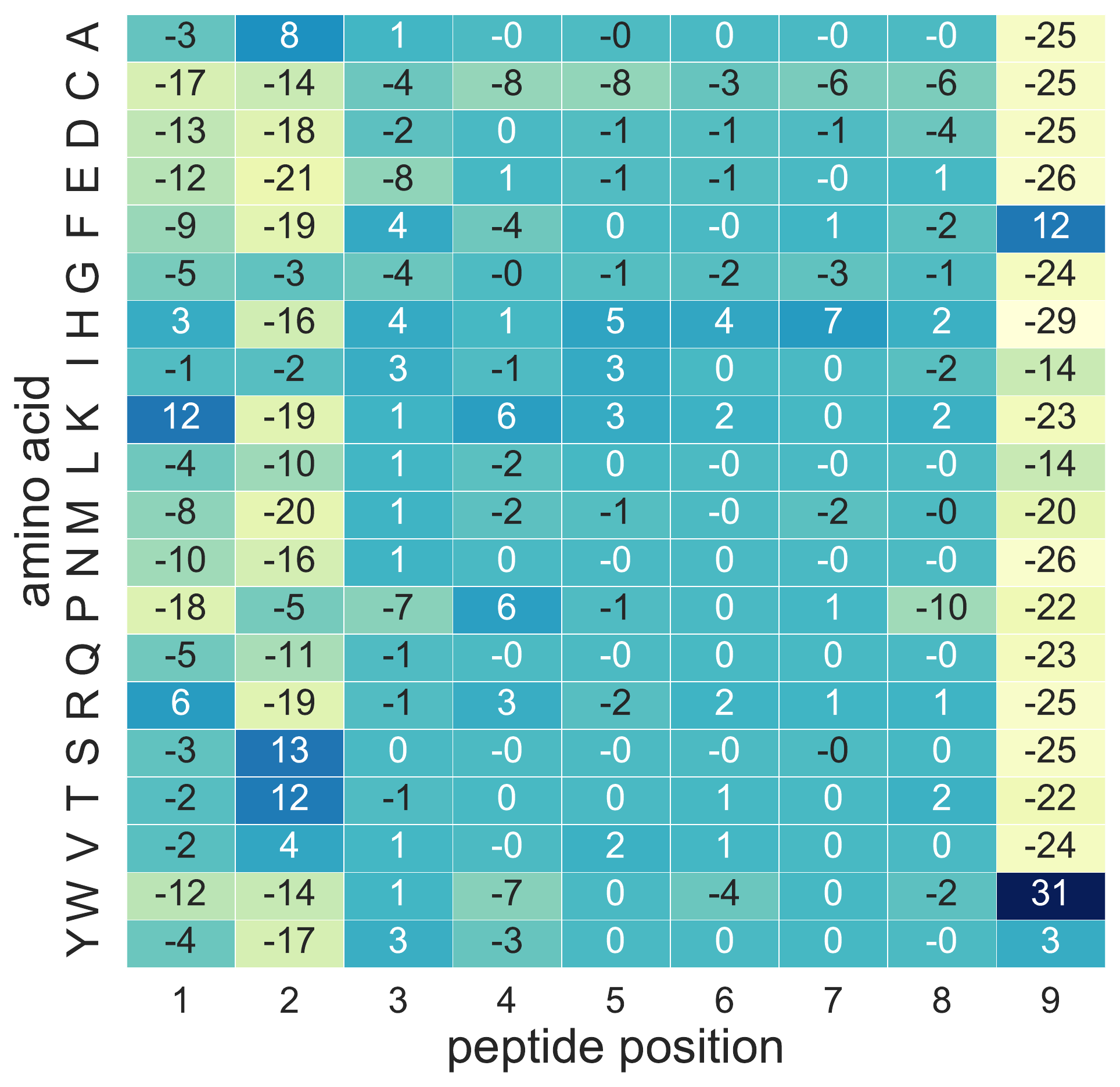}
					\end{center}
				\end{minipage}
				
				\noindent \begin{figure}[H] {\centering
						\includegraphics[width=11cm]{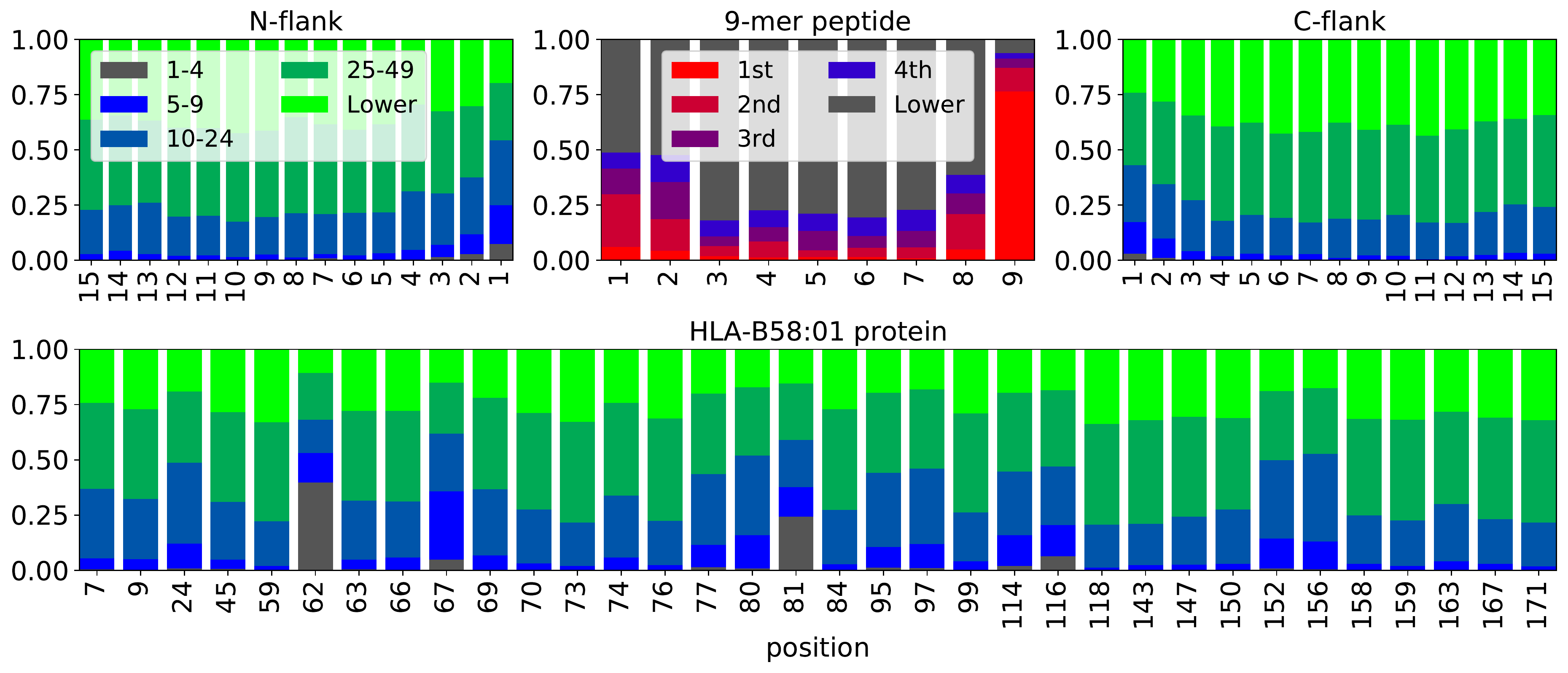} 
						\captionof{figure}{Motifs (top left), mean \gls{SHAP} values (top right) and \gls{LIME} feature ranks}
						\label{fig:LIME_\HLAallele} }
				\end{figure}

		\renewcommand{\HLAallele}{HLA-B5802}
		\renewcommand{\HLAalleleName}{HLA-B*58:02}
		\subsection{\HLAalleleName}
				\noindent \begin{minipage}[c][4.2cm][t]{7cm}
					\begin{center}
					\includegraphics[width=4cm]{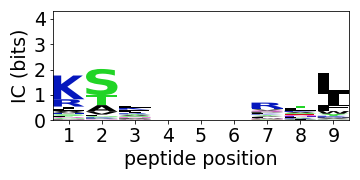}
					\includegraphics[width=4cm]{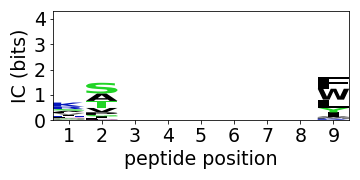} 
					\end{center}
				\end{minipage}
				\begin{minipage}[c][4.2cm][t]{7cm}
					\begin{center}
					\includegraphics[width=4.5cm]{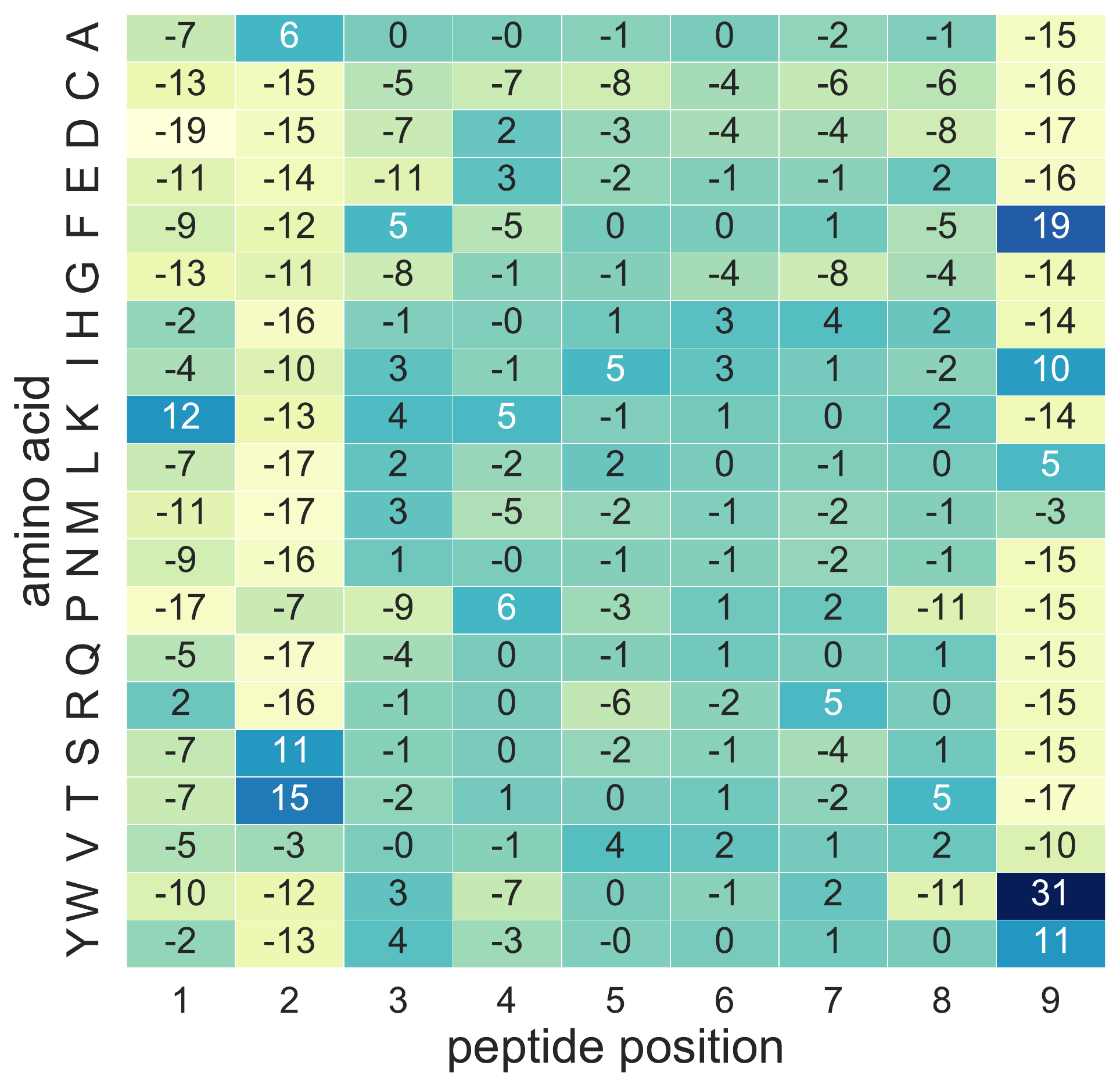}
					\end{center}
				\end{minipage}
				
				\noindent \begin{figure}[H] {\centering
						\includegraphics[width=11cm]{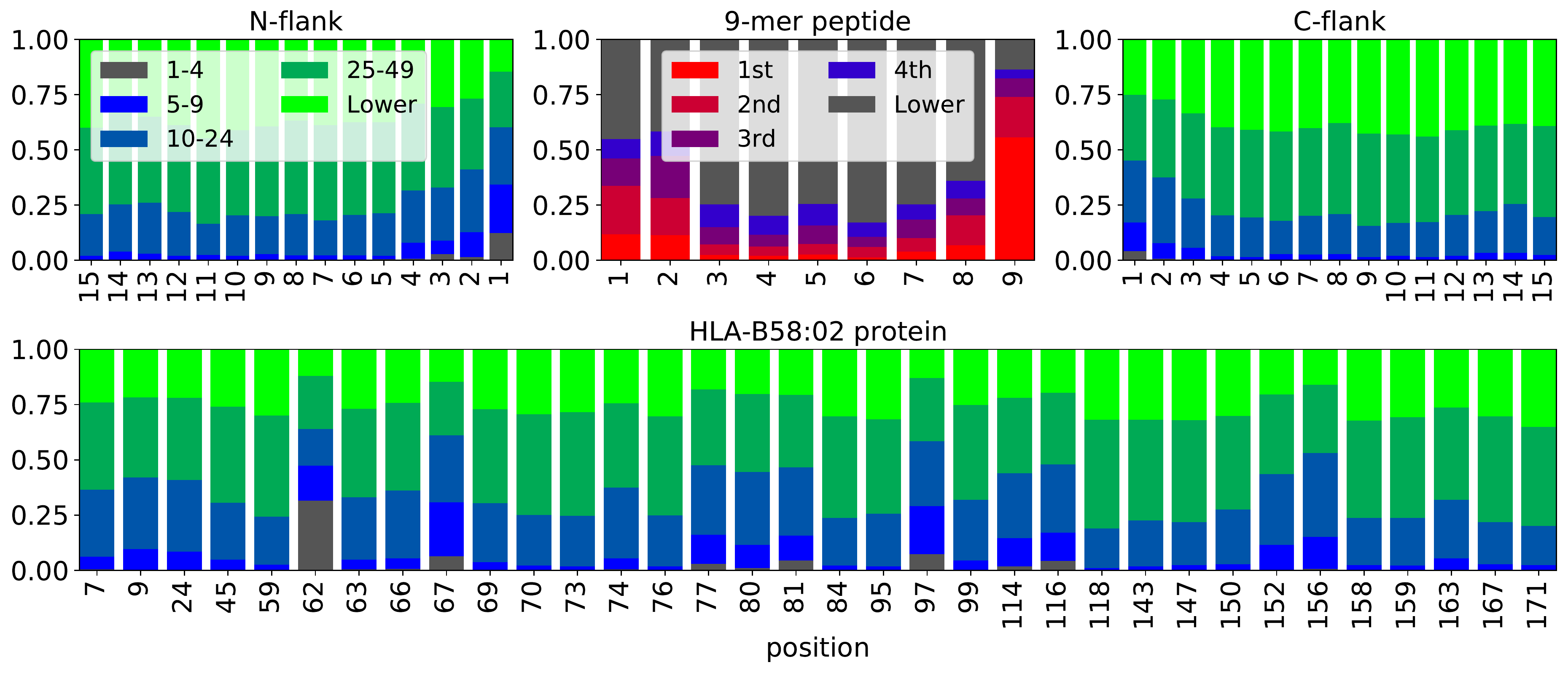} 
						\captionof{figure}{Motifs (top left), mean \gls{SHAP} values (top right) and \gls{LIME} feature ranks}
						\label{fig:LIME_\HLAallele} }
				\end{figure}

		\renewcommand{\HLAallele}{HLA-C1502}
		\renewcommand{\HLAalleleName}{HLA-C*15:02}
		\subsection{\HLAalleleName}
				\noindent \begin{minipage}[c][4.2cm][t]{7cm}
					\begin{center}
					\includegraphics[width=4cm]{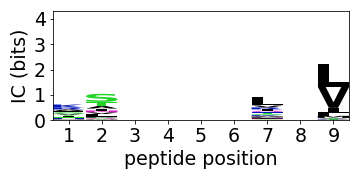}
					\includegraphics[width=4cm]{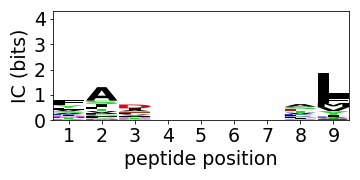} 
					\end{center}
				\end{minipage}
				\begin{minipage}[c][4.2cm][t]{7cm}
					\begin{center}
					\includegraphics[width=4.5cm]{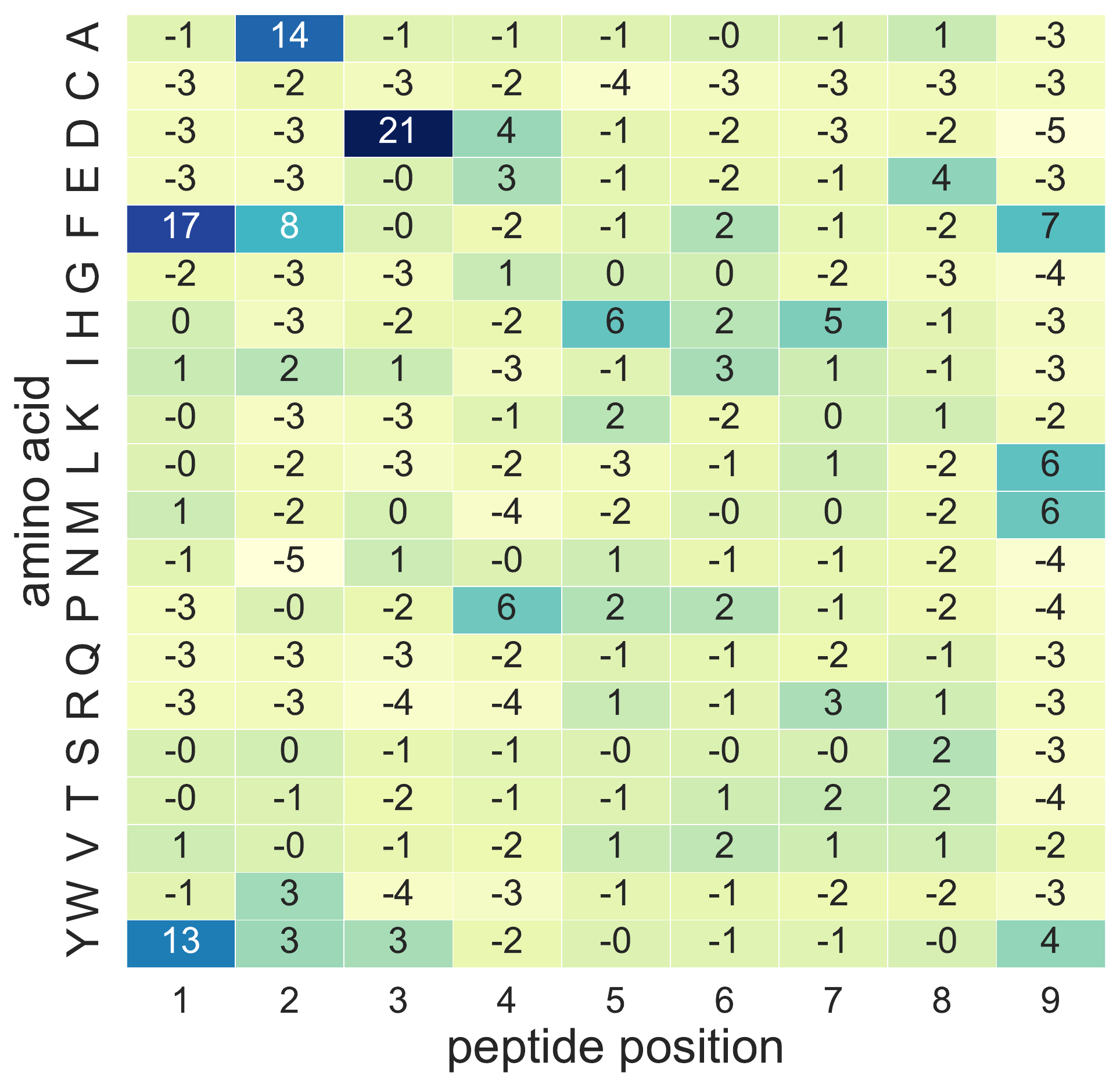}
					\end{center}
				\end{minipage}
				
				\noindent \begin{figure}[H] {\centering
						\includegraphics[width=11cm]{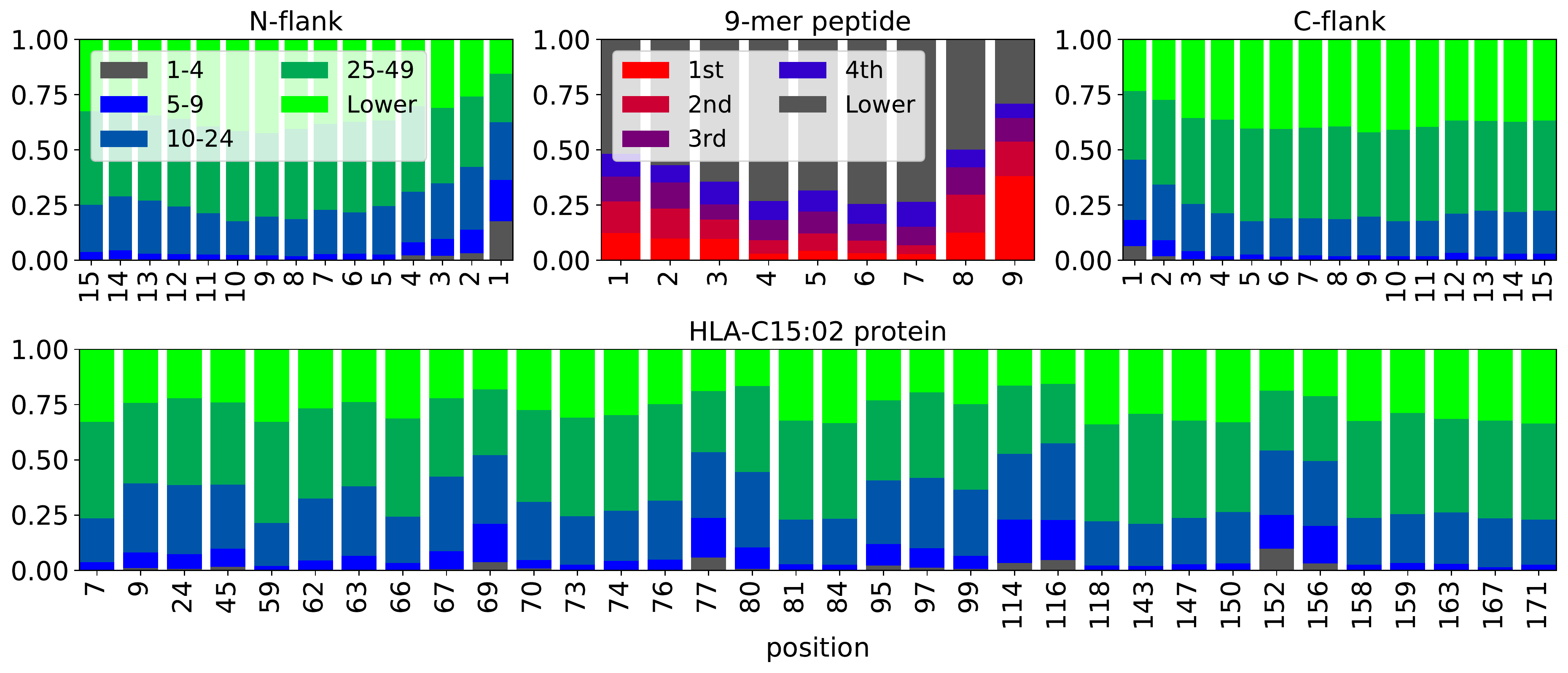} 
						\captionof{figure}{Motifs (top left), mean \gls{SHAP} values (top right) and \gls{LIME} feature ranks}
						\label{fig:LIME_\HLAallele} }
				\end{figure}

		\renewcommand{\HLAallele}{\HLAalleleIII}
		\renewcommand{\HLAalleleName}{\HLAalleleIIIname}
		\subsection{\HLAalleleName}
				\noindent \begin{minipage}[c][4.2cm][t]{7cm}
					\begin{center}
					\includegraphics[width=4cm]{\HLAallele_actual.png}
					\includegraphics[width=4cm]{\HLAallele_\finalModel.png} 
					\end{center}
				\end{minipage}
				\begin{minipage}[c][4.2cm][t]{7cm}
					\begin{center}
					\includegraphics[width=4.5cm]{shap_\HLAallele_\finalModel_all.pdf}
					\end{center}
				\end{minipage}
				
				\noindent \begin{figure}[H] {\centering
						\includegraphics[width=11cm]{lime_\HLAallele.pdf} 
						\captionof{figure}{Motifs (top left), mean \gls{SHAP} values (top right) and \gls{LIME} feature ranks}
						\label{fig:LIME_\HLAallele} }
				\end{figure}
				
		\renewcommand{\HLAallele}{HLA-C1701}
		\renewcommand{\HLAalleleName}{HLA-C*17:01}
		\subsection{\HLAalleleName}
				\noindent \begin{minipage}[c][4.2cm][t]{7cm}
					\begin{center}
					\includegraphics[width=4cm]{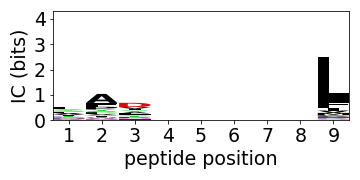}
					\includegraphics[width=4cm]{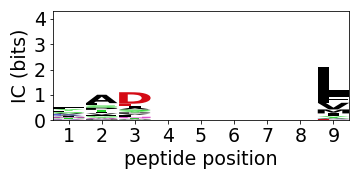} 
					\end{center}
				\end{minipage}
				\begin{minipage}[c][4.2cm][t]{7cm}
					\begin{center}
					\includegraphics[width=4.5cm]{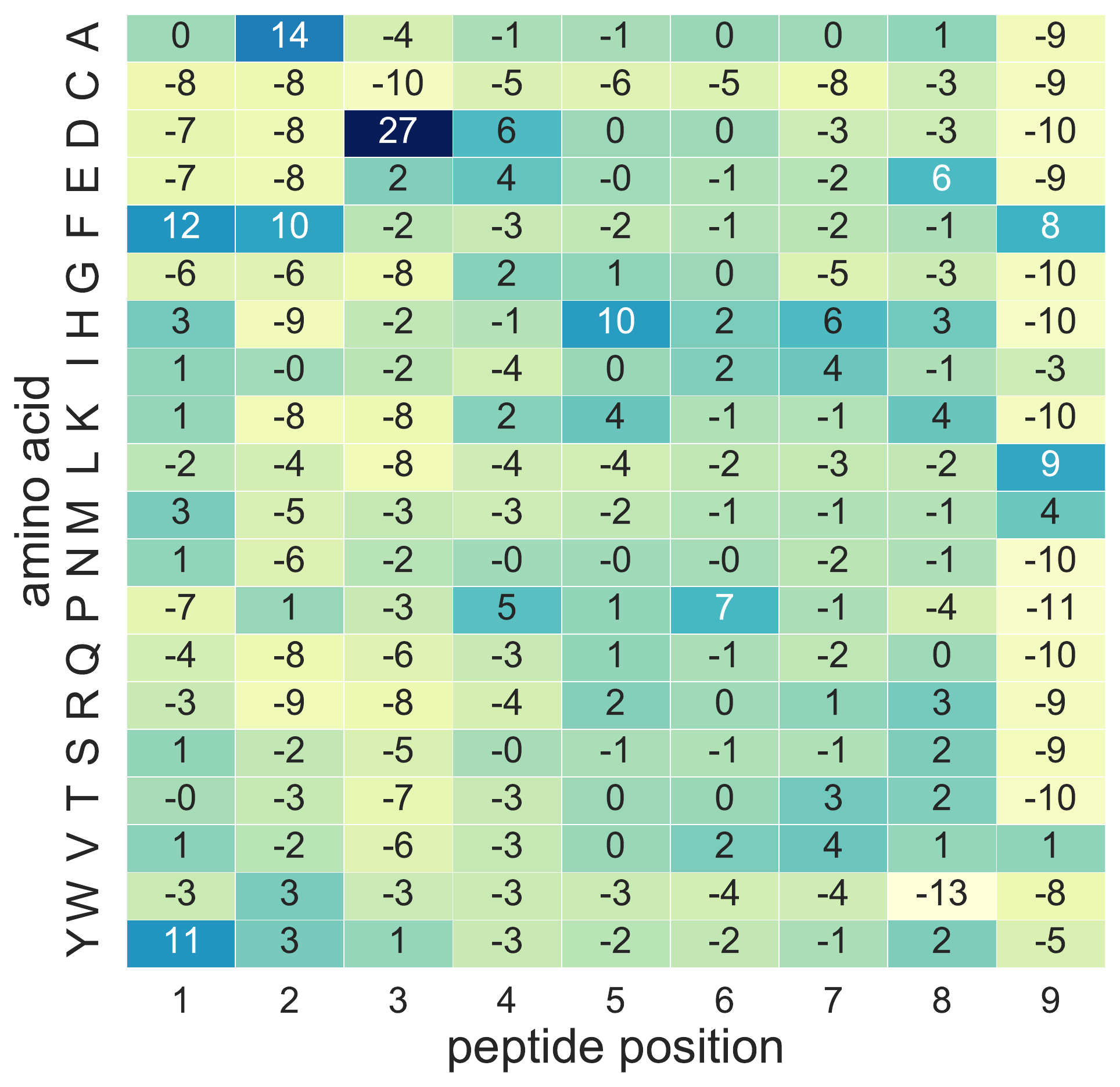}
					\end{center}
				\end{minipage}
				
				\noindent \begin{figure}[H] {\centering
						\includegraphics[width=11cm]{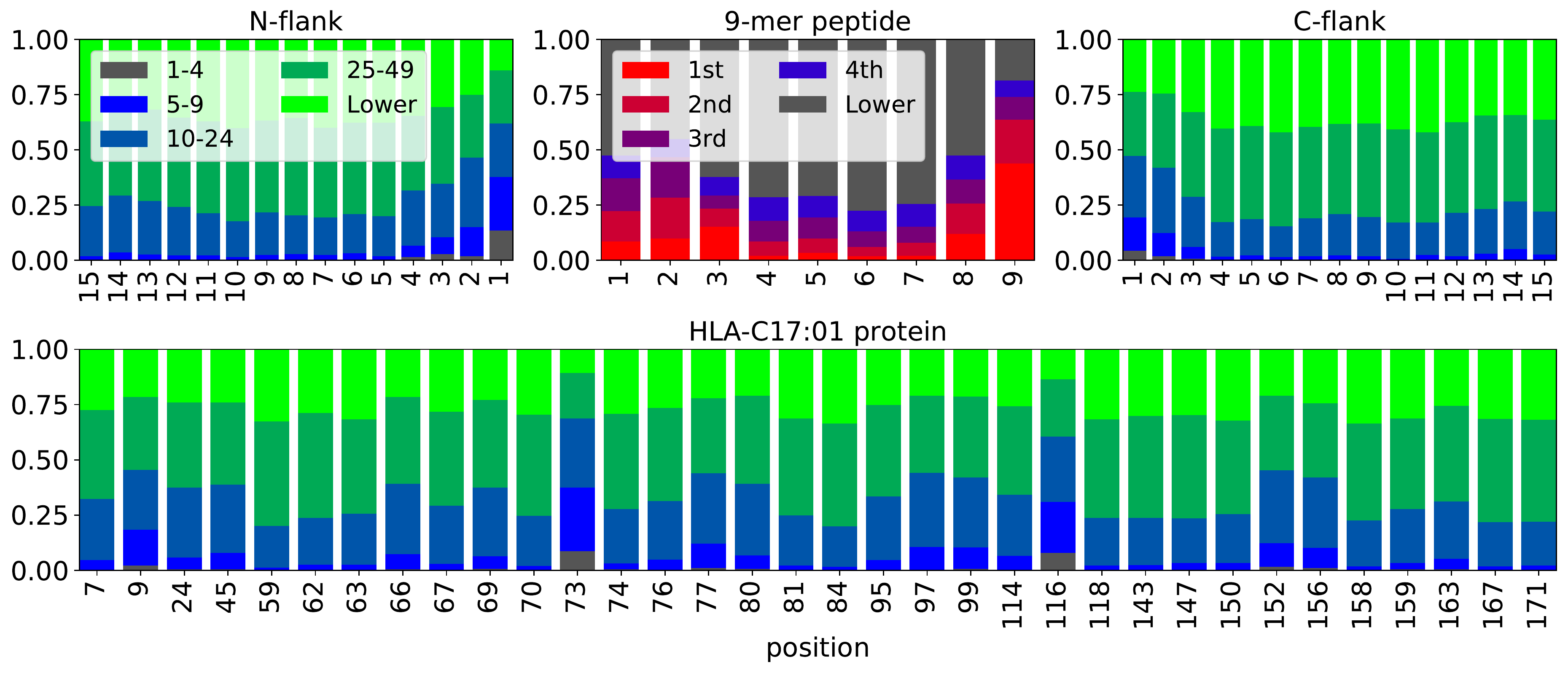} 
						\captionof{figure}{Motifs (top left), mean \gls{SHAP} values (top right) and \gls{LIME} feature ranks}
						\label{fig:LIME_\HLAallele} }
				\end{figure}

\end{document}